\documentclass[12pt]{report}

\usepackage[utf8]{inputenc}
\usepackage[a4paper, margin=1in ]{geometry}
\usepackage{amssymb}
\usepackage{amsmath}
\usepackage{graphicx}
\usepackage{float}
\usepackage{tabularx}
\usepackage[normalem]{ulem}
\usepackage{pdfpages}
\usepackage{lipsum}
\usepackage{tikz}
\usepackage[numbers]{natbib}
\usepackage{multicol}
\usepackage{setspace}
\usepackage[T1]{fontenc}
\usepackage[most]{tcolorbox}
\usepackage[pages=some]{background}
\usepackage{etoolbox}
\usepackage{forloop}
\usepackage{afterpage}

\usepackage{slashed}
\usepackage{bm}
\usepackage{braket}
\usepackage{amsthm}
\usepackage[compat=1.1.0]{tikz-feynman}
\usepackage{tikz}
\usepackage{caption}
\usepackage{subcaption}
\usepackage{algorithm2e}
\usepackage{fancyhdr}
\usepackage{color}
\usetikzlibrary{arrows}

\usepackage{titlesec}
\usepackage{xcoffins}

\newcommand\headerdisplay[1]{%
  \huge
  \vskip.5\baselineskip
  \filcenter\MakeUppercase{#1}%
  \vskip.0\baselineskip
}
\NewCoffin\mytmpa
\NewCoffin\mytmpb
\newcommand\placeabove[3][0pt]{%
 \SetHorizontalCoffin\mytmpa{#2}%
 \SetHorizontalCoffin\mytmpb{#3}%
 \JoinCoffins*\mytmpb[hc,t]\mytmpa[hc,b](0pt,#1)%
 \TypesetCoffin\mytmpb
}

\titleclass{\part}{top} 
\titleformat{\part}[frame]
  {\normalfont\vfil}
  {\filcenter\placeabove[2\baselineskip]{\Large PART}{\huge\enspace\thepart\enspace}}
  {1em}
  {\headerdisplay}[\vfil]
\titlespacing*{\part}{0pt}{0pt}{20pt}

\patchcmd{\chapter}{\thispagestyle{plain}}{\thispagestyle{chapterhead}}{}{}
\patchcmd{\part}{\thispagestyle{plain}}{\thispagestyle{empty}}{}{}

\usepackage[Glenn]{fncychap}

\definecolor{light}{rgb}{0.4, 0.4, 0.4}
\def\light#1{{\color{light}#1}}

\fancypagestyle{chapterhead}{%
    \fancyhf{}%
    \fancyfoot[C]{\thepage}%
}

\definecolor{cffffff}{RGB}{255,255,255}

\usepackage{tocloft}
\def\Title{Gribov Problem and Stochastic Quantization}

\def\Author{Adithya A Rao}

\def\InternalSupervisor{Dr.\ Vikash K. Ojha}

\def\ExternalSupervisor{Prof.\ Laurent Baulieu}

\def\ExternalSupervisorInsti{Sorbonne University}

\def\Year{2023 - 2024}

\title{\huge {{\Title}} \vspace{20pt}}
\author{\Author \\ Department of Physics\\ Sardar Vallabhbhai National Institute of Technology (SVNIT), Surat\\\vspace{30pt}\\ Supervised by\\
\ExternalSupervisor{}, \ExternalSupervisorInsti{}, Paris\\ \InternalSupervisor{}, SVNIT, Surat\vspace{230pt}} 
\date{Subimitted in partial fulfillment of the requirements for the degree of Master of Science in Department of Physics, SVNIT, Surat \\ \Year}

\backgroundsetup{
scale=1,
angle=0,
opacity=1,
contents={%
\hspace{0.05\paperwidth}\includegraphics[width=0.95\paperwidth,height=\paperheight]{assets/border.jpg}}%
}

\newcommand\exchange[2]{#2#1}%
\newcommand\includepatternloop[5]{%
  \include{#5#3#1#4}%
  \ifnum#1<\expandafter\exchange\expandafter{\number#2}{} %
     \exchange{\expandafter\includepatternloop\expandafter{\number\numexpr#1+1\relax}{#2}{#3}{#4}{#5}}%
  \fi
}%

\newcommand{\del}{\partial}
\newcommand{\pp}[2]{{\displaystyle\frac{\del #1}{\del #2}}}
\newcommand{\dd}[2]{{\displaystyle\frac{d #1}{d #2}}}
\newcommand{\DD}[2]{{\displaystyle\frac{\delta #1}{\delta #2}}}

\begin{document}

%%%%%%%%%%%%%%%%%%%%%%%%%%%%%%%%%%%%%%%%%%%%%%%%%%%%%%%%%%%%%%%%%%%%%%%%%%%%%

% cover page 
% \include{preamble/cover}

%%%%%%%%%%%%%%%%%%%%%%%%%%%%%%%%%%%%%%%%%%%%%%%%%%%%%%%%%%%%%%%%%%%%%%%%%%%%%

%preamble

\pagenumbering{gobble}
% \addtocontents{toc}{\protect \contentsline {chapter}{Declaration}{i}{}}
% \addtocontents{toc}{\protect \contentsline {chapter}{External Supervisor's Certificate}{ii}{}}
% \includepdf[pages=-]{preamble/external_certificate.pdf}
% \include{preamble/declaration}
% \addtocounter{page}{1}
% \includepdf[pages=-]{preamble/external_certificate.pdf}
% \addcontentsline{toc}{chapter}{}
% \include{preamble/certificate}
% \addtocontents{toc}{\protect \contentsline {chapter}{Certificate}{iii}{}}
% \includepdf[pages={6}]{preamble/scanned.pdf}
% \include{preamble/examiner_certificate}
% \addtocontents{toc}{\protect \contentsline {chapter}{Examiner's Certificate}{iv}{}}
% \includepdf[pages={6}]{preamble/scanned.pdf}
% \include{preamble/acknowledgement}
\maketitle 
%%%%%%%%%%%%%%%%%%%%%%%%%
% Created on 03-10-2023 %
%                       %
% Author: Adithya A Rao %
%%%%%%%%%%%%%%%%%%%%%%%%%

\newpage

% \addcontentsline{toc}{chapter}{Abstract}

% \BgThisPage{}
\begin{center}
    \Large
    {
        \vspace{40pt}
        {
            \textbf{
                \uppercase{Abstract}
            }
        }
    }
\end{center}
\normalsize
\vspace{30pt}
% \onehalfspacing

The standard procedure for quantizing gauge fields is the Faddeev-Popov quantization, which performs gauge fixing in the path integral formulation and introduces additional ghost fields. This approach provides the foundation for calculations in quantum Yang-Mills theory. However, in 1978, Vladimir Gribov showed that the gauge-fixing procedure was incomplete, with residual gauge copies (called Gribov copies) still entering the path integral even after gauge fixing. These copies impact the infrared behavior of the theory and modify gauge-dependent quantities, such as gluon and ghost propagators, as they represent redundant integrations over gauge-equivalent configurations. Furthermore, their existence breaks down the Faddeev-Popov prescription at a fundamental level. To partially resolve this, Gribov proposed restricting the path integral to the Gribov region, which alters the gluon propagator semiclassically in a way that points to gluon confinement in the Yang-Mills theory.\\

In this thesis, we comprehensively study the Gribov problem analytically. After reviewing Faddeev-Popov quantization, the BRST symmetry of the complete Lagrangian and the Gribov problem in depth, we detail Gribov's semi-classical resolution involving restriction of the path integral to the Gribov region, outlining its effects on the theory. Further, we elucidate stochastic quantization prescription for quantizing the gauge fields. This alternate quantization prescription hints towards a formalism devoid of the Gribov problem, making it an interesting candidate for quantizing and studying the non-perturbative regime of gauge theories.\\

\newpage
\normalsize
\onehalfspacing
% \setstretch{1.5}
\newpage
% % \BgThisPage{}
\addtocontents{toc}{\protect\thispagestyle{plain}}
% \addcontentsline{toc}{chapter}{Contents}

% \pagenumbering{gobble}
\tableofcontents{}
%%%%%%%%%%%%%%%%%%%%%%%%%%%%%%%%%%%%%%%%%%%%%%%%%%%%%%%%%%%%%%%%%%%%%%%%%%%%%
% \blankpage{}

\newpage
% \thispagestyle{empty}
% \hspace{0pt}
% \vfill 
% \begin{center}
%   \doublespacing
%   % \huge {\Cancellaresca To my family and my friends who always believed in me and had my back, and to the one in whom my belief further deepened as I got deeper into physics.}
%   \huge {\Cancellaresca To Physics \dots}
%   \onehalfspacing
% \end{center} 
% \vfill 
% \hspace{0pt}
% \newpage
% \blankpage{}

\setcounter{page}{0}
\pagenumbering{arabic}

%%%%%%%%%%%%%%%%%%%%%%%%%%%%%%%%%%%%%%%%%%%%%%%%%%%%%%%%%%%%%%%%%%%%%%%%%%%%%

% Body
\pagestyle{fancy}
% \setboolean{@twoside}{true} 
%%%%%%%%%%%%%%%%%%%%%%%%%
% Created on 03-10-2023 %
%                       %
% Author: Adithya A Rao %
%%%%%%%%%%%%%%%%%%%%%%%%%

\newpage

\part{The Gribov Problem}
% \afterpage{\blankpage}

\chapter{Prelude - Gauge Field Theories}
\vspace{-35pt}

% Quantum gauge field theories, more generally called \textit{the Standard Model}, form the backbone of modern physics. These theories are mathematically rich but complex to the point that they have evaded our attempts at completely understanding them since they were first described. Except for the \(U(1)\) gauge theory which is Quantum Electrodynamics, all other gauge theories have non-commutating fields, leading to self-interaction terms in the Lagrangian that give rise to complicated dynamics. This makes the theory 
% \setcounter{section}{-1}
\section{Introduction}
% Quantum Chromodynamics is a quantum gauge field theory describing the interactions mediated by the strong force - one of the four fundamental forces in nature.  Except for the \(U(1)\) gauge theory which is Quantum Electrodynamics, all other gauge theories have non-commutating gauge fields, leading to self-interaction terms in the Lagrangian that give rise to complicated dynamics. 
% One such peculiar outcome is the fact that in Quantum Chromodynamics, at low energies the coupling constant becomes so large that one cannot apply the perturbative techniques to extract insights into the dynamics of the theory. At these energies, no free quark or gluon exists, and any observable state is a bound state, and we call the theory \textit{confined}. \\

% The concept of gauge invariance was introduced by Hermann Weyl to describe electrodynamics in analogy to Einstein's theory of gravity, with local gauge freedom corresponding to Einstein's coordinate invariance, and the vector potential \(A_{\mu}\) playing the same role as the Christoffel's symbols \(\Gamma^{\nu}_{\lambda \mu}\). 

The quantization of non-Abelian gauge theories, such as quantum chromodynamics (QCD), is a fundamental challenge in theoretical physics. QCD is the theory that describes the strong interaction, one of the four fundamental forces of nature, and governs the behavior of quarks and gluons, the fundamental constituents of nucleons.\\

While QCD has been successful in explaining many phenomena in the perturbative regime, where the coupling strength is small, its non-perturbative aspects, which dominate at low energies and large distances, remain a significant challenge. This is because the coupling strength in QCD decreases with an increase in energy, and therefore at low energies the perturbative techniques which rely on writing an asymptotic series in the coupling constant break down. One of the key non-perturbative features of QCD is confinement, which refers to the fact that at low energies, quarks and gluons are never observed as free particles but are always confined within nucleons. This phenomenon is believed to be related to the non-Abelian nature of QCD and the self-interactions of the gluon field. Although there is no experimental evidence for the confinement of gluons, one expects them to be confined owing to them being colorful, and the gluon confinement is expected to be purely a quantum effect.\\

One of the problems in the quantization of non-Abelian gauge theories like QCD is the Gribov problem, which arises due to the redundant degrees of freedom present even after performing the gauge fixing. This redundancy leads to an overcounting of equivalent field configurations, resulting in an ambiguity in the path integral measure and the propagators of the theory.
The Gribov problem was first identified by Vladimir Gribov in the quantization of Yang-Mills theories in the Coulomb gauge. He discovered that the Coulomb gauge condition does not completely fix the gauge freedom, leading to the existence of multiple gauge copies, known as Gribov copies, which correspond to the same physical configuration but have different gauge field representations, all of them counted in the path integral.\\

This ambiguity in the gauge fixing procedure has far-reaching consequences for the non-perturbative aspects of gauge theories like QCD. The Gribov problem is believed to play a crucial role in the confinement mechanism and the dynamical generation of a gluon mass scale. Over the years, various approaches have been proposed to address the Gribov problem, including the Gribov-Zwanziger framework. However, these formulations often rely on certain approximations and may not fully capture the non-perturbative dynamics of the theory.\\

An alternative approach to addressing the Gribov problem is through stochastic quantization, a formalism introduced by G.\ Parisi and Yong-Shi Wu. Stochastic quantization reformulates the path integral of a quantum field theory as an equilibrium limit of a stochastic process, effectively replacing the time evolution of the system with a fictitious time evolution governed by a Langevin equation. \\

In this thesis, we will look at the origin and effects of the Gribov problem and also present the stochastic quantization formulation as an alternative quantization prescription that has the potential to circumvent some of the issues associated with the Gribov problem and provide insights into the non-perturbative aspects of gauge theories like QCD

\section{Gauge Freedom And Yang-Mills Theories}

The concept of gauge invariance was introduced by Hermann Weyl to describe electrodynamics in analogy to Einstein's theory of gravity, with local gauge freedom corresponding to Einstein's coordinate invariance, and the vector potential \(A_{\mu}\) playing the same role as the Christoffel's symbols \(\Gamma^{\nu}_{\lambda \mu}\). \\

To see exactly how the gauge fields arise, consider a theory with \(N\) spinor fields, \(\psi(x) \equiv \psi_\alpha^a(x)\) where \(\alpha\) denotes the Dirac indices, and \(a = 1,\dots, N\). This theory associates each spacetime point \(x\) with an \(N\)-dimensional vector space \(\Psi_x\) in which the spinors are defined.\ \textit{Gauge} refers to the basis for each of the vector spaces \(\Psi_x\). The physical content of the theory is required to be independent of the choice of basis of these vector spaces, and this freedom in choice of basis is called the gauge freedom. \\

The naive Dirac Lagrangian for the spinor fields in this theory can be written as 
\begin{equation}
    \mathcal{L} = \bar{\psi}(x) (i\slashed{\del} - m )\psi(x)
\end{equation}
This Lagrangian is \textit{naive} because it does not account for the freedom in the choice of basis for the spaces \(\Psi_x\). It contains the regular spacetime derivative \(\del_\mu\) which depends on comparing spinors at infinitesimally separated points, and since these live in different spaces spanned by different basis, there is no way to compare the spinors at two points, and consequently, the definition of derivative breaks down. \\

There is a simple way to circumvent this problem - choose one set of basis and enforce it on all the vector spaces at all points. But the derivative defined in this way is not gauge-invariant and is not something we are looking for. What we are looking for is an object that transforms a vector in one space into a vector in the space at an infinitesimally separated point. To see how we can define such an object, first, let us write the spinor fields in terms of the basis at each point.

\begin{equation}
    \psi_\alpha(x) = \psi^a_\alpha(x) ~u_a(x) 
\end{equation}
where \(u_a\) forms a basis for the space \(\Psi_x\). \\
Now when going from \(x\) to \(x+dx\) let the actual value of the spinor change by an amount 
\begin{equation}
    d\psi  = \psi(x+dx) - \psi(x)
\end{equation}
The \(\del_\mu \psi\) contains the contributions from both the actual change in the value of the spinor and the change in the orientation of the basis. Therefore 
\begin{equation}
    d\psi  = (\del_\mu\psi_a) dx^\mu~u_a + \psi_a du_a 
\end{equation}
Suppose in going from \(x\) to \(x+dx\), the basis of the internal spaces rotates by \(d\theta^k = (\del_\mu \theta^k)dx^\mu\) which is a set of \(N_g\) angles. Since the rotation group on \(N\) dimensional complex vectors is \(SU(N)\), \(N_g\) is the number of generators for \(SU(N)\). The rotation operator, which would be an element of \(SU(N)\) is therefore given by
\begin{equation}
    U(dx) = \exp \left( -iq ~d\theta^k~T^k_{ab}  \right)
\end{equation}
where \(T^k_{ab}\) are the generators of the \(SU(N)\) group.\\

This rotates the basis as 
\begin{equation}
    U(dx) ~u_a = \exp \left( -iq ~d\theta^k~T^k_{ab}  \right)u_b
\end{equation}
Expanding this upto first order in \(dx\), we get
\begin{equation}
    u_a + du_a = \left(\delta_{ab} - iq ~(\del_\mu \theta^k)dx^\mu~T^k_{ab}  \right)u_b
\end{equation}
which gives 
\begin{equation}
    du_a = - iq ~(\del_\mu \theta^k)~T^k_{ab}u_bdx^\mu
\end{equation}
If we call \((\del_\mu \theta^k(x))T^k_{ab} = (A_\mu)_{ab}\) the connection, then the total change in \(\psi\) is given as 
\begin{equation}
    d\psi  = \left( (\del_\mu\psi_a)\delta_{ab} - iq (A_\mu)_{ab} \psi_a\right) u_b dx^\mu
\end{equation}
From this, we can write the gauge-covariant derivative operator defined as called as the covariant derivative as  
\begin{equation}
    D_\mu \psi_a = (\del_\mu \delta_{ab} - iq (A_\mu)_{ab} )\psi_b
\end{equation}

The comparator has values as a function of spacetime points and therefore defines a field on its own which we call the \textit{gauge fields}, and since we require that the covariant derivative transform \textit{covariantly} under gauge transformations, we see that the gauge fields must transform as 
\begin{equation}
    A_{\mu}(x)\rightarrow \exp(i\theta(x)^kT^k)A_\mu(x)\exp(-i\theta(x)^kT^k)  -\frac{i}{q}(\del_\mu\exp(i\theta(x)^kT)^k) ~\exp(-i\theta(x)^kT^k)
\end{equation}
Infinitesimally, this transformation looks like 
\begin{equation}
    A_\mu^k \rightarrow A_\mu^k - \frac{1}{q}\del_\mu \alpha^k + f^{klm} A^l_\mu \alpha^m = A_\mu^k - \frac{1}{q}D_\mu^{\text{(adj)}~kl}\alpha^l
\end{equation}
where \(f^{klm}\) are the structure constants for the \(SU(N)\) group, and \(D^{(\text{adj})}\) denotes the action of covariant derivative in the adjoint representation.\\

With this, we can write a gauge invariant Dirac Lagrangian density as 
\begin{equation}
    \mathcal{L}_{Dirac} = \psi(x) (\slashed{D} - m )\psi(x)
\end{equation}

To make the vector field \(A\) dynamic, we add Lorentz covariant and gauge invariant terms, quadratic in derivatives and fields. Noting that the covariant derivative transforms covariantly and that the commutator \([D_\mu,~ D_\nu]\) has terms consisting of only \(A\) and its derivatives, we can construct a Lagrangian density as 
\begin{equation}
    \mathcal{L_{A}} = \frac{1}{2}\frac{1}{ig} Tr[D_\mu, D_\nu]^2
\end{equation}

Calling \(\displaystyle \frac{1}{ig} [D_\mu, D_\nu]_{ab} = F_{\mu\nu~ab} = F_{\mu\nu}^k T^k_{ab}\) and using the normalization for the trace of generators of the group, we write 

\begin{equation}
    \mathcal{L_A} = \frac{1}{4}F_{\mu\nu}^k F^{\mu\nu}_k
    \label{eq:yangmillsLagrangian}
\end{equation}

The field theory with a non-abelian gauge group \(SU(N)\) and a Lagrangian density of the gauge fields specified by the equation (\ref{eq:yangmillsLagrangian}) is called a \textit{Yang-Mills theory}.\\

\(F_{\mu\nu}^k\), which is also called the field strength tensor, can be written in terms of \(A\) as 
\begin{equation}
    F_{\mu\nu}^k = \del_\mu A_\nu^k - \del_\nu A_\mu^k - g f^{klm}A_\mu^l A_\nu^m
\end{equation}

The Lagrangian density constructed as prescribed contains the terms cubic and quartic in \(A\) and its derivatives. These higher-order terms arise purely due to the non-abelian nature of the gauge group, and the existence of these terms implies that the field \(A\) itself is a charged field, and thus there is a multitude of self-interaction vertices in even the simplest Feynman diagrams which is the reason for the beautiful complexity and richness of the theory.

\newpage

%%%%%%%%%%%%%%%%%%%%%%%%%
% Created on 03-10-2023 %
%                       %
% Author: Adithya A Rao %
%%%%%%%%%%%%%%%%%%%%%%%%%

\newpage
\chapter{Quantizing the Gauge Fields}
\vspace{-45pt}

In this chapter, we provide a comprehensive introduction to the Faddeev-Popov quantization procedure which forms the basis for all quantum field theory calculations. Developed by Ludvig Faddeev and Victor Popov in the 1960s, this method addresses the issue of redundant degrees of freedom present in gauge field theories. In the context of QCD, the gauge symmetry leads to an overcounting of physically equivalent configurations, which can result in inconsistencies and ambiguities in the quantization process.
The Faddeev-Popov procedure introduces auxiliary fields, known as ghost fields, to cancel out the unphysical degrees of freedom associated with the gauge redundancy. The ghost fields, while unphysical themselves, play a crucial role in ensuring the gauge invariance of the quantized theory and maintaining the correct counting of physical degrees of freedom.

\section{The Failure of the Traditional Quantization}
Given a field theory in field \(A\) and a classic action functional \(S[A]\) defined on it, the traditional quantization method is to take the classic action functional and write a path integral for it defined as
\begin{equation}
    Z[J] = \int_\mathcal{C} \mathcal{D}A~ \mathrm{e}^{\left(-S[A] - \int  d^4x J\cdot A\right)}
    \label{eq:fullPathIntegralNoGaugeFixing}
\end{equation}
where \(J\) is a function of spacetime and not fields, and is introduced as a mathematical tool for generating Green's functions. The integral over the configuration space \(C\) with the (ill-defined) measure \(\mathcal{D}A\) refers to integration over all possible field configurations, and to make the measure well-defined, we further assume suitable regularisations.\\

The Yang-Mills Lagrangian inherently contains terms cubic and quartic in fields, and therefore it does not describe a free theory but rather describes an interacting theory. To quantize Yang-Mills perturbatively, we consider only the quadratic part of the Lagrangian, and take the self-interaction terms as perturbations. In such a setting, the free path integral for the Yang-Mills is written as 
\begin{equation}
    \begin{split}
        Z_{quad}[J] &= \int_\mathcal{C} \mathcal{D}A~\exp{ \left( -\int  d^4x \frac{1}{4}(\del_\mu A_\nu^k - \del_\nu A_\mu^k)^2 + \int d^4x~J_\mu^k A_\mu^k \right)}\\
        & = \int_\mathcal{C} \mathcal{D}A~\exp{ \left(  -\int d^4 x \frac{1}{2}A_\nu^k(\del_\mu\del_\nu - \delta_{\mu\nu}\del^2 )A_\mu^k + \int  d^4x~J_\mu^k A_\mu^k \right) }
    \end{split}
\end{equation}
One can solve the Gaussian integral to obtain 
\begin{equation}
    Z_{quad}[J] = \frac{1}{\det(G_{\mu\nu})}\exp\left( \frac{1}{2}\int  d^4x ~J_\mu ^k~ G^{-1}_{\mu\nu}~J_\nu^k \right)
    \label{eq:NoGaugeFixingZ}
\end{equation}
where \(G_{\mu\nu} = \del_\mu\del_\nu - \delta_{\mu\nu}\del^2 \).\\
In a usual quantum field theory, \(G^{-1}_{\mu\nu}\) would be called the propagator, and by making use of Wick's theorem one writes all the correlation functions in terms of the above-defined propagator. One could do the same in the case of Yang-Mills too, if the inverse of \(G_{\mu\nu}\) existed. But there exist zero modes of the form \(\del_\mu \lambda\) for which \((\del_\mu\del_\nu - \delta_{\mu\nu}\del^2)\del_\mu \lambda = 0\), making the operator non-invertible and therefore making \(Z_{quad}[J]\) ill-defined. This is a consequence of the gauge freedom, which makes the operator \(\del_\mu\del_\nu - \delta_{\mu\nu}\del^2\) a many-to-one map, breaking its invertibility. \\

Another shortcoming of this method can be seen as follows. The path integral is formally an integration over all possible field configurations, and gauge freedom means that all the gauge equivalent configurations are counted multiple times too, meaning that the path integral gives rise to an infinity, which is the size of the gauge transformation group. This makes the path integral infinite and therefore lose its meaning.\\

To consistently define the theory, we need to eliminate the gauge degree of freedom.  One way to do this would be to introduce a constraint on the space of configurations that is satisfied by only one out of the infinite gauge copies. A simple constraint would be of the form \(\del_\mu A_\mu = \sqrt{\alpha}~b\). \\

In the Abelian gauge theory (QED), we proceed by promoting this constraint to an equation of motion, and we write a modified Lagrangian of the form 
\begin{equation}
    \mathcal{L}_{QED} = \frac{1}{4}(\del_\mu A_\nu - \del_\nu A_\mu)^2  + \frac{\alpha}{2}b^2 +  b\del_\mu A_\mu
\end{equation}

The equation of motion from this Lagrangian implies that \(\del A\) follows the free field equation, \(\square \del A= 0\). This means that we can freely assign to it the field \(b\) and this does not modify the physical content of the theory. Introducing this term breaks gauge invariance and hence makes the quadratic operator invertible.\\

In the case of Yang-Mills, one cannot follow such a gauge fixing procedure due to the presence of the self-interaction terms. The equation of motion for \(\del A\) does not follow the free field equation, which means that assigning to it the field \(b\) naively truly modifies the physical content of the theory. Therefore the gauge fixing procedure that made QED consistent fails in the case of Yang-Mills. 

\section{Faddeev-Popov Ghosts}
% Gauge invariance implies that the path integral, eq (\ref{eq:fullPathIntegralNoGaugeFixing}), sums over equivalent configurations. Thus, the integral gets multiplied by the volume of the symmetry group, which in the case of Yang-Mills is \(\infty\) since the symmetry group is \(\mathcal{G}^\infty\). Thus, to get the right path integral, we need to factor out the volume from the integral. \\

Faddeev and Popov, in their 1967 work, suggested a way to quantize gauge fields without performing gauge fixing in the Lagrangian. They did the quantization by introducing a delta function in gauge constraint directly into the path integral. This would mean that the path integral counts only those configurations that satisfy the gauge condition and discards those that do not. \\
Consider the identity in the form
\begin{equation}
    \int \mathcal{D}\mathcal{F}(^GA)~~\delta(\mathcal{F}(^GA)-\lambda) = 1
    \label{eq:gaugeFixingIdentity}
\end{equation}

where \(\mathcal{F}(^GA) - \lambda = 0\) is the gauge condition, with \(^GA\) referring to the gauge transformed field.\\

We can change the integration measure from \(\mathcal{DF}(^GA)\) to \(\mathcal{D}G\) 
\begin{equation}
    \int \mathcal{D}G~\biggr | \det\left( \DD{\mathcal{F}(^GA)}{G} \right) \biggr | ~\delta(\mathcal{F}(^GA)-\lambda) = 1
\end{equation}
with the determinant arising as the Jacobian of the variable change. \\

Since varying over \(\lambda\) simply gives a class of gauge fixing conditions that give rise to the same physical theory, we can integrate over \(\lambda\) with a Gaussian measure to obtain

\begin{equation}
    \int \mathcal{D}\lambda \exp{\left(-\int d^4x \frac{\lambda^2}{2\alpha}\right)} \int  \mathcal{D}G~ \delta(\mathcal{F}(^GA)-\lambda) ~\biggr |  \det\left( \frac{\delta \mathcal{F}(^GA)}{\delta G} \right) \biggr |  = C(\alpha)
\end{equation}

This identity can be inserted into the path integral eq (\ref{eq:fullPathIntegralNoGaugeFixing})
\begin{equation}
    \begin{split}
        Z[J] &= \mathcal{N} \int  \mathcal{D}G\int_\mathcal{C}\mathcal{D}A \biggr |  \det\left( \frac{\delta \mathcal{F}(^GA)}{\delta g} \right) \biggr | \exp{\left(\displaystyle  -S[A] -\int d^4x \left( \frac{\mathcal{F}(^GA)^2}{2\alpha} + J.A  \right)\right)}
    \end{split}
\end{equation}
Since \(S[A]\) and \(\mathcal{D}A\) are gauge invariant, we can perform two consecutive changes of variables \(A \rightarrow ^GA\) and then \(^GA \rightarrow A\) without altering the integration, giving
\begin{equation}
    \begin{split}
        Z[J] = \mathcal{N}\int \mathcal{D}G\int_\mathcal{C}\mathcal{D}A & \biggr |  \det\left( \frac{\delta \mathcal{F}(^GA)}{\delta G} \right)\biggr | \biggr |_{G=1} ~ \\
        &\times \exp\left(  -S[A] -\int d^4x  \left( \frac{\mathcal{F}(A)^2}{2\alpha} - J.A  \right)\right)
    \end{split}
\end{equation}

Now, since the integrand is independent of \(G\), the integral \(\displaystyle \int \mathcal{D}G\) simply gives the volume of the gauge group factored out of the path integral and therefore can be absorbed into the normalization without altering the values of observables.\\

In this thesis, we will mainly deal with the gauge fixing condition of the form \(\del_\mu A_\mu = 0\). This means the operator inside the determinant is of the form (since infinitesimal gauge transformation is of the form \(A_\mu^k \rightarrow A_\mu^k - D_\mu ^k \alpha \))
\begin{equation}
    \det\left(\DD{\del_\mu (A_\mu^k (x) - D_\mu^{\text{(adj)}~kl}\alpha^l(x))}{\alpha^l(y)} \right)\biggr |_{\alpha=0} = \det(-\del_\mu D_\mu^{(\text{adj})~kl} \delta(x-y))
\end{equation}

Using the property of Grassman integrals, we write this determinant as a path integral 
\begin{equation}
    \begin{split}
        \det(-\del_\mu &D_\mu^{(\text{adj})~kl} \delta(x-y)) =\\
        & \int \mathcal{D}\Omega~\mathcal{D}\bar\Omega~ \exp\left(-\int d^4x d^4y~\bar{\Omega}^k(x)(\del_\mu D_\mu^{(\text{adj})~kl} \delta(x-y)) \Omega^l(y)\right)        
    \end{split}
\end{equation}
where \(\Omega\) and \(\bar\Omega\) are Grassman-valued scalars that transform in the same representation as \(A\). \\

With this introduction, we can write the complete Lagrangian of a consistent Yang-Mills theory as 
\begin{equation}
    \mathcal{L} = \frac{1}{4}(F_{\mu\nu}^k)^2 + \frac{1}{2\alpha}(\del_\mu A_\mu^k)^2 + \bar{\Omega^k}\del_\mu D_\mu^{(\text{adj})~kl} \Omega^l
    \label{eq:fullLagrangian}
\end{equation}

The fields \(\Omega\ \&~ \bar{\Omega}\) satisfy the Klein-Gordon equation but are also anticommuting fields. Since this is a violation of the spin-statistics theorem, the excitations of these fields can not be physical. These fields can only enter the equations as internal lines and not asymptotic states. Thus due to their non-physical nature, they are called as \textit{Faddeev-Popov ghosts}.

\subsection{Feynman Rules For The Complete Yang-Mills}
\label{sec:FeynRules}
From the Lagrangian (\ref{eq:fullLagrangian}), we can write down the free propagators and the interaction vertices for the gauge and ghost fields as follows

\begin{itemize}
    \item \textbf{Free Gauge Propagator}: \\
    The quadratic part of the Lagrangian in gauge fields is \(\displaystyle \frac{1}{4}(\del_\mu A_\nu^k - \del_\nu A_\mu^k)^2 + \frac{1}{2\alpha}(\del_\mu A_\mu^k)^2 \). Thus the free gauge propagator would be the inverse of the operator \(\displaystyle \del_\mu \del_\nu - \delta_{\mu\nu}\del^2 - \frac{1}{\alpha}\del_\mu\del_\nu\). In momentum space, the free propagator is
    \begin{equation}
        \braket{A_\mu^{k}(p) A_\nu^{l}(-p)} = \raisebox{-1.6ex}{\feynmandiagram [horizontal=a to b] { 
            a [particle=$A^k_\mu$] -- [gluon, momentum=\(p_\mu\)] b [particle=$A^l_\nu$],
           };} = \frac{-i\delta_{kl}}{p^2 + i\epsilon}\left( \delta_{\mu\nu}- (1-\alpha)\frac{p_\mu p_\nu}{p^2}  \right)
    \end{equation}
    \item \textbf{Free Ghost Propagator}: \\
    The free part of the ghost Lagrangian is simply that of the complex scalar field theory, and so the free ghost propagator in momentum space would be 
    \begin{equation}
        \braket{\bar{\Omega}^k(p)\Omega^l(-p)} = \raisebox{-1ex}{\feynmandiagram [horizontal=a to b] { 
            a [particle=$\bar{\Omega^k}$] -- [ghost, momentum=\(p_\mu\)] b [particle=$\Omega^l$],
        };} = \frac{-i\delta_{kl}}{p^2 + i\epsilon}
    \end{equation}
    \item \textbf{Gauge-Gauge Vertices}: \\
    There exist two gauge-gauge vertices, one for the interaction term \(g A_\nu^k f^{klm}A_\mu^l \del_\mu A_\nu^m\) with the vertex factor \(g p_\mu f^{klm}\) with \(p_\mu\) being the momentum of \(A_\nu^m\), and another with interaction term \(g^2f^{klm}A_\mu^l A_\nu^m f^{kno}A_\mu^n A_\nu^o \) with the vertex factor \(g^2 f^{klm}f^{kno}\)
    
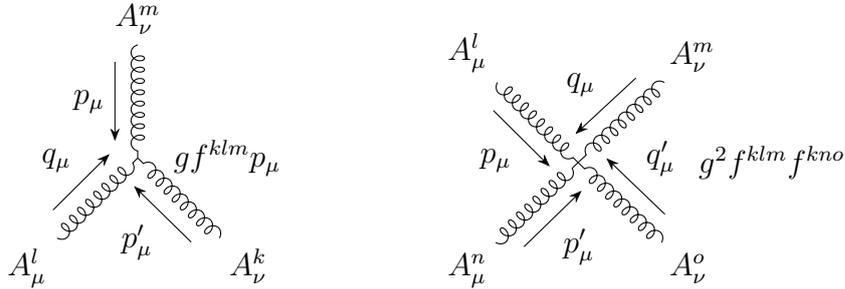
\begin{center}
        \begin{tikzpicture}
            \begin{feynman}
                \vertex (a) ;
                \vertex[above =of a] (i1) {\(A_\nu^m\)};
                \vertex[below left =of a] (i2) {\(A_\mu^l\)};
                \vertex[below right =of a] (i3) {\(A_\nu^k\)};
                \vertex[right=0.8em  of a] {\(gf^{klm} p_\mu\)};
                \diagram* {
                    (i1) -- [gluon, momentum'=$p_\mu$] (a),
                    (i2) -- [gluon, momentum=$q_\mu$] (a),
                    (i3) -- [gluon, momentum=$p'_\mu$] (a) ,
                    };
            \end{feynman}
        \end{tikzpicture}~~~~~~~~~~~~~~\begin{tikzpicture}
            \begin{feynman}
                \vertex (a) ;
                \vertex[above left=of a] (i1) {\(A_\mu^l\)};
                \vertex[above right=of a] (i2) {\(A_\nu^m\)};
                \vertex[below left =of a] (i3) {\(A_\mu^n\)};
                \vertex[below right =of a] (i4) {\(A_\nu^o\)};
                \vertex[right=3.5em of a] {\(g^2f^{klm}f^{kno}\)};
                \diagram* {
                    (i1) -- [gluon, momentum'=$p_\mu$] (a),
                    (i2) -- [gluon, momentum'=$q_\mu$] (a),
                    (i3) -- [gluon, momentum'=$p'_\mu$] (a) ,
                    (i4) -- [gluon, momentum'=$q'_\mu$] (a) ,
                    };
            \end{feynman}
        \end{tikzpicture}
    \end{center}

    \item \textbf{Gauge-Ghost Vertex}: \\
    There also exists a gauge-ghost vertex corresponding to the interaction term \(\bar{\Omega}^kf^{klm}A_{\mu}^{l}\del_\mu \Omega^m\) with the vertex factor being \(gf^{klm}p_\mu\), where \(p_\mu\) is the momentum of \(\Omega^m\)
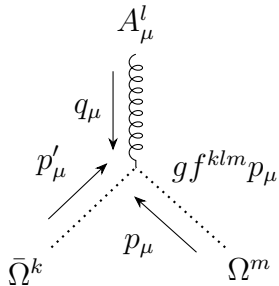
    \begin{center}
        \begin{tikzpicture}
            \begin{feynman}
                \vertex (a);
                \vertex[above =of a] (i1) {\(A_\mu^l\)};
                \vertex[below left =of a] (i2) {\(\bar{\Omega}^k\)};
                \vertex[below right =of a] (i3) {\(\Omega^m\)};
                \vertex[right=0.8em  of a] {\(gf^{klm} p_\mu\)};
                \diagram* {
                    (i1) -- [gluon, momentum'=$q_\mu$] (a),
                    (i2) -- [ghost, momentum=$p'_\mu$] (a),
                    (i3) -- [ghost, momentum=$p_\mu$] (a) ,
                    };
            \end{feynman}
        \end{tikzpicture}
    \end{center}
    \item \textbf{Negative Sign of Ghost Loops}: \\
    Since the ghost fields are anticommuting, the Feynman diagrams with ghost loops get multiplied by \(-1\).
\end{itemize}
These Feynman rules consistently define Yang-Mills theory encapsulating both gauge and ghost fields.\\

\section{Remarks On The Faddeev-Popov Method}
In the previous discussion, we overlooked a few crucial considerations, which seem trivial but are very important for the construction of the theory.\\

In the identity, eq (\ref{eq:gaugeFixingIdentity}), the change of integration variable should be done as 
\begin{equation}
\displaystyle \int \mathcal{D}G ~\delta(\mathcal{F}(^GA)-\lambda) = \prod\sum  \frac{1}{\biggr|\DD{\mathcal{F}(^GA)}{G}\biggr |}
\label{eq:gaugeFixingIdentityFull}
\end{equation}
where the summation stands over all the zeros of the gauge fixing functional. We made an assumption here that the gauge fixing condition is satisfied by a single configuration out of all gauge equivalent configurations. This reduces the ugly expression in the right-hand side of the above, to a simple product of the absolute value of the eigenvalues of the Faddeev-Popov operator \(\DD{\mathcal{F}(^GA)}{G}\). This can be then written as the determinant of the said operator. \\
If the gauge fixing condition is satisfied by multiple equivalent configurations, this procedure will break down.\\

Another assumption we made was of the positive definiteness of the Faddeev-Popov operator, which allowed us to ignore the absolute value requirement of the determinant. With only this assumption, it was possible to introduce the ghost fields and the Grassman integral. If the Faddeev-Popov operator has negative modes, then we cannot drop the absolute value operation, and therefore we cannot introduce the ghost fields. 
Further, if the FP operator has zero modes, then for that specific configuration, the determinant is zero, and hence the contribution of that configuration to the path integral is nullified.\\

It is these very considerations that later come back to bite us which will be discussed later in our later discussion of the Gribov problem. 

\section{The BRST Symmetry}

Let us take a detour and discuss the implications that the Faddeev-Popov quantization has for the theory. Since the introduction of the gauge fixing term and ghost fields breaks the gauge symmetry, we look for a new symmetry that encompasses all these fields, also encapsulating the residual gauge symmetry of the Lagrangian. This symmetry, called the BRST symmetry, was discovered by Carlo Maria Becchi, Alain Rouet, Raymond Stora, and Igor Viktorovich Tyutin, as a complete symmetry of the Faddeev-Popov effective action. \\

The ghost Lagrangian is invariant under the scaling of (anti)ghost fields as 
\begin{equation}
    \begin{split}
        \Omega &\rightarrow e^{\gamma}\Omega\\
        \bar\Omega &\rightarrow e^{-\gamma}\bar\Omega
    \end{split}
\end{equation}
where \(\gamma\) is a real, spacetime-independent parameter.\\
Noether's theorem implies a conserved current corresponding to this symmetry, which we will call the ghost number. We can compute the ghost number for the fields, \(\Omega\) carries a ghost number \(+1\) while \(\bar{\Omega}\) carries a ghost number \(-1\). Since \(A_\mu\) has no (anti)ghost components, it has ghost number 0.\\

We adopt here a matrix notation in adjoint representation, where we saturate spacetime indices of fields (if present) with \(dx^\mu\). For example, \(A = A_\mu^k T^k dx^\mu\). This is a matrix in adjoint representation with elements being one-forms in spacetime. The elements of this matrix are anticommuting under exterior product. Similarly \(\Omega = \Omega^k T^k\), the elements of which are Grassmann numbers, and hence are anticommuting. \\
Along with this, we introduce a universal graded bracket, \([X, Y] = X\pm Y\)
where the + sign occurs only in the case \(X\) and \(Y\) are both anticommuting.\\

It is immediately obvious that the matrix elements anticommute if its form degree is odd and the ghost number is even or the other way round. The matrix elements are always commuting when both the form degree and ghost number are odd or even. 
Thus we can assign a generalized grading, which defines if the matrix elements are commuting or anticommuting. This grading is defined as the sum of the form degree and the ghost number. If the grading is even, then the matrix elements commute. Otherwise, they anticommute. \\

Here we can introduce a structure, what we will call a generalized n-form, \((\Omega_n)^a_b,~a+b=n\) whose ghost number is \(a\) and form degree is \(b\). The (anti)commutative nature of such a structure depends on n (mod 2), similar to their n-form counterparts.\\
In the space of these structures, there must exist an operator \(\hat{d}\), which behaves like the exterior derivative that acts on the space of n-forms. The operator \(\hat{d}\) must increase the grading by 1, which can be achieved in two ways. The first way is to increase the form degree by 1, while the second way is to increase the ghost number by 1. That is, we require the action of \(\hat{d}\) on \((\Omega_n)^a_b\) as \(\hat{d}(\Omega_n)^a_b = (\Omega_{n+1})^{a}_{b+1} + (\Omega_{n+1})^{a+1}_b\).\\
There already exists the regular exterior derivative \(d\) that increases the form degree by 1, and hence we demand a new operator \(s\), which we will call the BRST operator, that increases the ghost number by 1. With this we would have \(\hat{d} = d+s\), with both \(d\) and \(s\) being operators of grading 1.\\

The operator \(\hat{d}\) constructs an anti-commuting structure from a commuting one and vice versa, and hence it must hold that \(\hat{d}^2 = d^2 + ds + sd + s^2 = 0\). Since \(s\) and \(d\) act on orthogonal spaces, \([d,s] = ds + sd = 0\) should hold. Thus, we arrive at the condition \(s^2 = 0\), i.e., the BRST operator should be nilpotent of order 2. This condition, along with \(sd + ds = 0 \) defines how the operator \(s\) acts on the fields. \\

To see where such a geometrical structure arises from, we see that if we consider spacetime to be spanned by \(\{x\}\) and the gauge group to be spanned by \(\{y\}\), then the spinor field at any point in the extended space \((x,y)\) can be written as 
\begin{equation}
    \tilde{\psi}_a(\mathbf{x}, \mathbf{y}) = \exp\left(i y^k T^k_{ab}\right)\psi_b(\mathbf{x})
\end{equation}

The local connections on the extended space would be the generalized one form
\begin{equation}
    \tilde{A}^k = A^k_\mu dx^\mu + \Omega^k_m dy^m
\end{equation}
where the first term is the regular gauge field, and the second term, which is the local connection in the gauge space, can be recognized with the ghost fields.\\
In this extended space, ghosts are not some artifacts that need to be introduced to the theory to make it consistent but rather are the fundamental objects of the gauge theory. \\

The BRST operator arises in this model simply as the exterior derivative in the \(y\) space. 
\begin{equation}
    s = dy^m \del_m
\end{equation}
The deduction that \(s^2 = 0\) from the previous discussion is now straightforward since \(s\) is the exterior derivative on \(y\) space, which means it should be nilpotent. This operator can be, in a sense, interpreted as a generator of translations in the gauge space. Gauge invariance implies invariance under translations in the gauge space, which means that we expect the Lagrangian to be invariant under such translations. To check this, we proceed as follows. \\

From the conditions \(s^2=0\) and \([d,s]=0\), it is pretty straightforward to construct the action of \(s\) on the fields
\begin{equation}
    \begin{split}
        sA &= \frac{1}{g}d\Omega + [A,\Omega] = D^\text{(adj)}\Omega\\
        s\Omega &= \frac{1}{2g}[\Omega,\Omega]\\
        s\bar\Omega &= -\frac{1}{g\alpha}(\del_\mu A_\mu)
    \end{split}
\end{equation}
These equations can also be constructed from the ``horizontality'' condition on the generalized curvature \(\tilde F\), which states that for a physical theory, the only non-zero components of the curvature tensor should be the \(dx_\mu \wedge dx_\nu\) components. The above equations are precisely the equations that state that the other components of \(\tilde F\) are zero.\\

As expected, we can notice from the above equations that the action of the operator \(\eta s\) on the gauge fields is exactly that of an infinitesimal gauge transformation, if we recognize \(\eta \Omega\) with the field of gauge parameters, where \(\eta\) is an infinitesimal g-valued constant introduced to make the gauge transformations boson valued.\\

It can be seen that the Lagrangian eq (\ref{eq:fullLagrangian}) is indeed invariant under these transformations of the fields, which means that the BRST operator generates a symmetry group of the Lagrangian.\\

A very important consideration that we overlooked here is that for this Lagrangian,  \(s^2 \bar\Omega \propto \del D^{\text{(adj)}}\Omega\) is \(0\) only when the EOM for the ghost field, \(\del D^{\text{(adj)}}\Omega = 0\) is imposed by hand. This means that the current Lagrangian is in an on-shell representation of the BRST algebra.\\

To get a representation that closes off-shell too, we re-introduce the non-dynamical auxiliary bosonic field \(b\) of mass dimensions 2, whose equations of motion give the gauge fixing condition. Such a Lagrangian would be 
\begin{equation}
    \mathcal{L} = \frac{1}{2} F^2 + \frac{\alpha}{2}b^2 + b \del A + \bar\Omega \del D^\text{(adj)} \Omega
\end{equation}
Because the mass dimension of \(b\) is 2, it can never enter the Lagrangian dynamically, and hence it behaves as a background field. But at the same time, its introduction makes the BRST operator closed off-shell hence completing its definition.\\ 

With the introduction of the auxiliary field, the action of the BRST operator becomes
\begin{equation}
    \begin{split}
        sA &= \frac{1}{g}dA +[A,\Omega]\\
        s\Omega &= \frac{1}{2g}[\Omega,\Omega]\\
        s\bar\Omega &= b\\
        sb &= 0
    \end{split}
\end{equation}
We immediately see that these transformations close off-shell too. Thus, the BRST symmetry forms a symmetry of the complete Lagrangian. \\

Many authors have suggested that rather than arriving at the BRST symmetry as the symmetry group of the Lagrangian, the right thing to do would be to consider a symmetry group with nilpotent generators and construct a Lagrangian that is invariant under the transformations generated by this group, stating the BRST invariance to be the first principle from which a gauge theory should be constructed.\\

Nevertheless, BRST symmetry is not simply some mathematical curiosity, rather it plays a vital role in the renormalisability of the theory. Imposing the condition that the physical states of the theory should be annihilated by \(s\) removes all the unphysical states of the theory from the physical subspace, and also ensures the unitarity of the theory \cite{slavnovPhysicalUnitarityBRST1989}.
Some authors have also suggested that the violation of BRST symmetry due to the Gribov problem might also lead to confinement effects.

\section{Non-Trivial Structure Of The Gauge Group}
So far, we have not considered how the gauge group is structured, since it was irrelevant to the discussion. But to address the Gribov problem, it is necessary to understand the notion of \textit{small} and \textit{large} gauge transformations and a discussion of the structure of the gauge group becomes necessary here. \\

The gauge group, in its simple sense, defined as \(\mathcal{G}^\infty\) is trivial, considering that all the gauge transformations can be smoothly connected to the identity. But this is not true actually, since we impose some restrictions on the allowed gauge transformations, thus making the structure non-trivial.\\

The conserved Noether charge associated with the gauge symmetry is called the color charge. One can construct the color charge operator from Noether's theorem as \cite{ildertonColourCopiesConfinement2007}
\begin{equation}
    Q\ket{\text{physical}} =\frac{1}{g}\int\!\mathrm{d}^{3}x~\partial_{i}E_{i}(x)\ket{\text{physical}}  =\frac{1}{g}\,\lim_{R\to\infty}\int_{S_{R}^{2}}\!\mathrm{d}s\cdot E(x)\ket{\text{physical}} 
\end{equation}
where \(E(x)\) is the gauge field equivalent of the QED electric field.\\

Since we require that the color charge operator remain invariant under gauge transformations, the following expression must hold,
\begin{equation}
    Q^{U}=\frac{1}{g}\lim_{R\to\infty}\int_{S_{R}^{2}}\mathrm{d}s\cdot U^{-1}(x){ E}(x){ U}(x) = Q
\end{equation}
Thus at \(R\rightarrow \infty\), we require that the gauge group element must behave as \(\displaystyle \lim_{x\rightarrow\infty} U(x) = U_\infty\) where \(U_{\infty}\) should be a constant that commutes with all other members of the group.\\

% This restriction is of the form 
% \begin{equation}
%     U(x)|_{x\rightarrow \infty} = c
% \end{equation}
% that is, approaching infinity from any direction we should get a constant transformation, usually taken to be identity. \\

% Many authors have pondered over this restriction, there have been a lot of reasonings as to why this is a viable condition, but no concrete reason exists as to why this restriction is necessary. In many places, this is taken as an axiom of the theory, and further consideration is done. \\
% One of the compelling arguments is that the Yang-Mills theory is a theory in which the sole effect of the field \(A_0\) on the theory is via the boundary values since the bulk values become irrelevant due to the gauge invariance, which means that some type of boundary condition should be imposed on the theory to make it sensible. \\
% Other arguments include that the physical potential should fall off as \(1/r\), and also 

Adding this restriction, it turns out that the allowed gauge transformations form disjoint subsets of the original gauge group, each characterized by a winding number and with elements smoothly connected to some paradigm. With the introduction of this non-trivial structure, we can talk about small and large gauge transformations. A small gauge transformation would be the one that would be smoothly connected to identity, while large ones would be the one that would belong to other homotopy classes, disconnected from identity. 
This disconnection is because any function that smoothly connects gauge transformations that belong to two different homotopy classes should pass through regions where the boundary condition is not satisfied. In such case, the gauge transformation does not remain a pure gauge transformation, and hence any smooth function between two homotopy classes must leave the domain of the allowed gauge transformations somewhere in between. \\

This property is reflected in the vacuum structure of the Yang-Mills theory. There are multiple vacuums of the Yang-Mills, each with a different topological charge, and these vacuums are not connected via small gauge transformations. The wavefunction can pick a global phase factor parameterized by a vacuum angle \(\theta\) when the system transitions from one vacuum to another, and this phase factor is physical observable. \\
The CP violation of the theory implies that the phase factor should have some value, but so far the experiments have constrained the value of the phase angle to be \(< 10^{-10}\). The relative absence of such a phase factor is termed the strong CP problem, which is beyond the scope of our discussion.

%%%%%%%%%%%%%%%%%%%%%%%%%
% Created on 03-10-2023 %
%                       %
% Author: Adithya A Rao %
%%%%%%%%%%%%%%%%%%%%%%%%%

\newpage
\chapter{The Haunting of the Gribov Copies}
\vspace{-45pt}
% TODO: Write short intro
In this chapter, we will discuss the Gribov problem in detail with explicit examples. The Gribov problem, first identified by Vladimir Gribov in the late 1970s, arises from the ambiguity in the gauge-fixing procedure, where multiple gauge field configurations, known as Gribov copies, can satisfy the same gauge condition while representing the same physical state. We also outline Gribov's original proposal for resolving this problem and describe the implications of implementing the Gribov restriction in a semiclassical approximation. Specifically, we will show how the restriction modifies the standard Faddeev-Popov approach and leads to the generation of a mass parameter for gluons.

\section{The Gribov Problem}
The ideal gauge fixing condition would be the one that will be satisfied by only one of the gauge copies for a given configuration. Mathematically, the existence of the ideal gauge fixing condition translates to a statement regarding the existence of a global section of the principle bundle of the theory. We can write a base manifold as the set of all configurations of the gauge fields \(\{A\}\). For each \(A \in \{A\}\), we assign a set of all equivalent configurations \(\{^G A\}\) called the gauge orbit. 
The orbits form a fiber bundle over the base manifold, the entire structure being called the principle bundle. The gauge fixing function, which picks representatives from each gauge orbit thus defines a continuous section map from the base manifold to the fiber bundle.\\

The ideal gauge condition that we demanded, which requires a section map to intersect the gauge orbit exactly once per orbit, translates to it being a global section. According to a theorem in topology, a principle bundle will admit a global section if and only if it is a trivial bundle. 
If the principle bundle is non-trivial, then no matter how ingenious the gauge fixing condition is devised to be, there will be multiple intersections with at least a few of the orbits, and the gauge fixing method will fail. Such a problem, if exists, would be termed the \textit{Gribov Problem}.\\

At this point, one might ask what are the consequences of the failure of gauge fixing. First of all, the elevation of the right-hand side of eq (\ref{eq:gaugeFixingIdentityFull}) to the inverse of the determinant would not be possible since the primary assumption in doing so was that the gauge condition is satisfied only once per orbit. 
Secondly, the existence of multiple gauge copies satisfying the gauge fixing condition implies the existence of \(\alpha\) such that \(\del_\mu D_\mu (\alpha) = 0\). This implies that the FP operator has zero modes, meaning that the determinant in the path integral becomes zero for some configurations, effectively causing their contribution to the integral to be nullified. \\
The existence of zero modes also implies that perturbations around these zero modes can create negative eigenvalues of the FP operator. We were able to introduce ghost fields by dropping the absolute value operation owing to the assumption that the Faddeev-Popov operator is always positive. The existence of these negative modes implies that the introduction of the Grassmann integral and the ghost fields become meaningless. \\

Thus, the non-uniqueness of gauge fixing conditions would lead to profound problems in the very foundations of the quantum Yang-Mills theory.

\section{Does The Gribov Problem Exist? }
Consider an explicit example of \(SU(2)\) Yang-Mills in the Coulomb gauge, \(A^0 =0,~\del_i A_i = 0\).\\

To simplify our discussion we consider one specific field configuration, the vacuum configuration in which the field curvature vanishes implying that the fields are in pure gauge \(A_i = U^{-1}\del_i U\). \\
Following Gribov, we take spherically symmetric gauge transformations of the form 
\begin{equation}
    U(r) = \exp\left( i\frac{1}{2} \alpha(r) \sigma_in_i  \right) = \cos\left( \frac{1}{2}\alpha(r)\right) + n_i\sigma_i \sin\left( \frac{1}{2}\alpha(r)\right)
\end{equation}
where \(n_i\) is the unit vector \(x_i/r\), and \(\sigma_i\) are the generators of \(SU(2)\)\\

With such a gauge transformation, the gauge copies of the vacuum would be

\begin{equation}
    \begin{split}
        A_i =& \left(\cos\left( \frac{1}{2}\alpha\right) - n_i\sigma_i \sin\left( \frac{1}{2}\alpha\right)\right) \\
        &\times \left( -\frac{1}{2} \sin\left( \frac{1}{2}\alpha\right) \dd{\alpha}{r} n_i - \frac{1}{r}\left( \delta_{ij} - n_i n_j \right)\sigma_j \sin\left( \frac{1}{2}\alpha\right) + n_j\sigma_j\frac{1}{2} \cos\left( \frac{1}{2}\alpha\right) \dd{\alpha}{r} n_i    \right)\\
        =& ~\frac{1}{2}\alpha' n_i (n_i\sigma_j) + \frac{1}{2r}\left( \delta_{ij} - n_i n_j \right)\sigma_j \sin\left( \alpha\right) + \sin^2\left(\frac{1}{2}\alpha \right)\frac{1}{r}\epsilon_{ijk}n_j \sigma_k
    \end{split}
\end{equation}

The Coulomb gauge condition in this case translates to the differential equation 
\begin{equation}
    r^2 \alpha'' + 2r\alpha' - 2 \sin \alpha = 0
\end{equation}
With a change of variable \(r = \exp(t)\), the equation becomes
\begin{equation}
    \alpha'' + \alpha' - 2\sin \alpha = 0
\end{equation}
This equation is an equation of motion for a damped pendulum, with \(\alpha\) being the angle from the unstable equilibrium of the pendulum, with a periodic external driving force, usually called the Gribov pendulum, with the pendulum initially being at \(\alpha=0\), the position of unstable equilibrium. \\
To avoid singularities at \(r = 0\), we require the condition \(\alpha(r = 0) = 0\) to be imposed. This condition translates to \(\displaystyle \lim_{t\rightarrow -\infty}\alpha(t) = \delta \exp(t)\). Following this, we can construct three different classes of solutions, \(\delta = 0,~\delta>0, ~\& ~\delta<0\). \\
The case \(\delta = 0\) gives the trivial vacuum \(\alpha(r) = 0\). The other two cases give topologically charged vacuum with charges \(\displaystyle \pm \frac{1}{2}\), with boundary conditions \(\displaystyle\lim_{t\rightarrow \infty}\alpha_{\pm\frac{1}{2}} = \pm\displaystyle\frac{\pi}{2}\), with \(\alpha(r) = \alpha_{\frac{1}{2}}(r)\) and \(\alpha(r) = - \alpha_{\frac{1}{2}}(r)\) respectively. \\

Therefore, the Coulomb gauge condition is satisfied by not one, but multiple vacuum configurations each belonging to one of the three homotopically distinct classes, i.e. separated by large gauge transformations
\begin{equation}
    \begin{split}
        A &= 0 \\
        A_i &= \exp\left(-i\alpha_{\frac{1}{2}}(r)  \sigma_in_i\right)~\del_i\exp\left(i\alpha_{\frac{1}{2}}(r)  \sigma_in_i\right)\\
        A_i &= \exp\left(i\alpha_{\frac{1}{2}}(r)  \sigma_in_i\right)~\del_i\exp\left(-i\alpha_{\frac{1}{2}}(r)  \sigma_in_i\right)
    \end{split}
\end{equation}
The three solutions that exist to the pendulum equation with the specified conditions are: the pendulum never falls (which is the trivial vacuum), the pendulum falls clockwise, and the pendulum falls counterclockwise.  
\vspace{10pt}

Gribov, in his seminal paper, considered a more general case of a general spherically symmetric gauge field configuration and showed that the gauge fixing condition fails in all cases of the gauge fields. Further, Singer proved that for any gauge theory and any covariant gauge condition, the gauge fixing procedure will always break down. \cite{singerRemarksGribovAmbiguity1978} \\

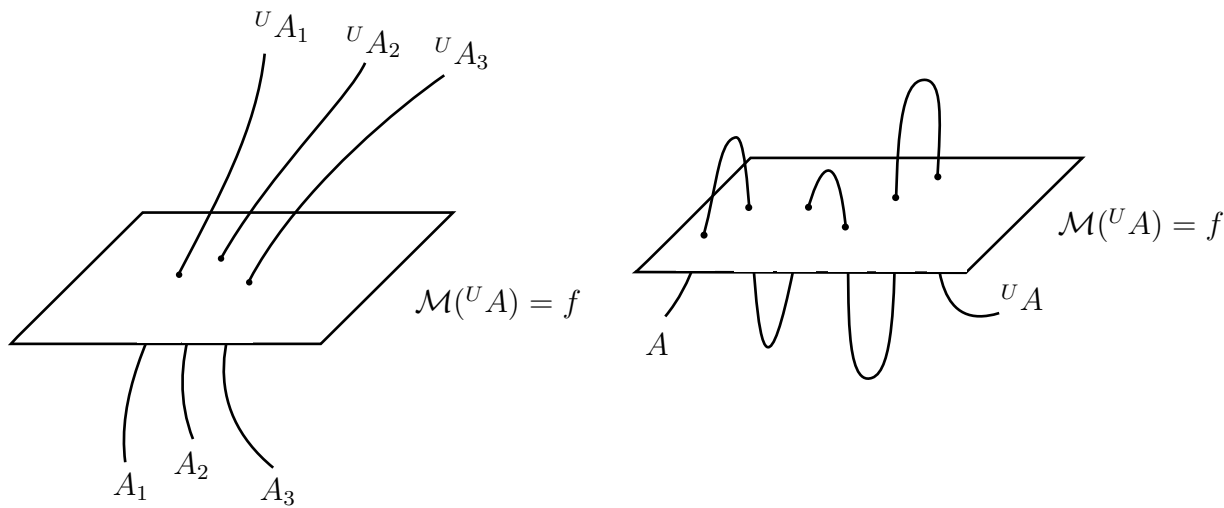
\begin{figure}[h]
    \begin{subfigure}[b]{0.48\textwidth}
        \centering
        \begin{tikzpicture}[y=1px, x=1px, yscale=0.3,xscale=0.3, inner sep=0pt, outer sep=0pt]
            \path[draw=black,line width=1px] (78.5358,479.3389) -- (234.2532,635.0564)
              -- (620.1617,635.0564) -- (455.6428,470.5375) -- (70.4114,470.5375) -- cycle;
            \path[draw=black,line width=1px] (212.5882,322.2674) .. controls (192.2021,
              485.1366) and (366.0469, 643.9116) .. (385.9084, 834.1039)(387.2625,601.2048);
            \path[draw=black,line width=1px] (296.7796,351.3092) .. controls (224.2426,
              538.3685) and (476.7559, 748.1245) .. (510.8040, 822.5905);
            \path[draw=black,line width=1px] (396.5257,315.4162) .. controls (213.7256,
              467.4461) and (485.0616, 719.6172) .. (609.1961, 807.0084);
            \path[fill=cffffff,line width=1px] (227.40074456692912, 557.7925417322833)
              rectangle (281.5633360629921, 472.4864655118109);
            \path[fill=cffffff,line width=1px] (283.13846551181103, 548.0843338582677)
              rectangle (328.6838437795276, 472.35293480314954);
            \path[fill=cffffff,line width=1px] (294.6280894488189, 577.2871181102362)
              rectangle (340.17346771653547, 501.55571905511806);
            \path[fill=cffffff,line width=1px] (325.01566866141735, 548.1134362204723)
              rectangle (370.5610469291339, 472.3820371653542);
            \path[fill=black,line width=1px] (279.5298, 557.1130) circle (4px);
            \path[fill=black,line width=1px] (331.6612, 577.4240) circle (4px);
            \path[fill=black,line width=1px] (366.6977, 547.6345) circle (4px);
            \path[fill=black,line width=1px] (197.2385,274.8417) node[above right]
              (text300){$A_1$};
            \path[fill=black,line width=1px] (379.8669,268.8884) node[above right]
              (text300-5){$A_3$};
            \path[fill=black,line width=1px] (272.6304,300.9637) node[above right]
              (text300-59){$A_2$};
            \path[fill=black,line width=1px] (373.4127,848.6838) node[above right]
              (text452){$^U A_1$};
            \path[fill=black,line width=1px] (597.0063,815.4426) node[above right]
              (text452-6){$^U A_3$};
            \path[fill=black,line width=1px] (487.3762,829.3259) node[above right]
              (text452-9){$^U A_2$};
            \path[fill=black,line width=1px] (572.5000,500.0197) node[above right]
              (text500){$\mathcal{M}(^U A) = f$};
          \end{tikzpicture}
          \caption{\light{Ideal case of gauge fixing where \(\mathcal{M}(^U A) = f\) is the gauge condition. Our assumption had it that each of the gauge orbits intersected the gauge fixing hyperplane only once. }}
    \end{subfigure}\hfill
    \begin{subfigure}{0.48\textwidth}
        \centering
        \begin{tikzpicture}[y=1px, x=1px, yscale=0.33,xscale=0.33, inner sep=0pt, outer sep=0pt]
            \path[draw=black,line width=1px] (104.2630,477.9849) .. controls (109.4978,
              481.0072) and (234.2532, 607.9751) .. (234.2532, 607.9751) --
              (609.3291,607.9751) -- (477.5401,477.1435) -- cycle;
            \path[draw=black,line width=1px] (137.8754,427.3977) .. controls (199.6155,
              502.3733) and (187.5712, 627.4561) .. (217.6795, 631.3292) .. controls
              (243.5416, 634.6560) and (228.8707, 214.3722) .. (282.6287, 480.9265) ..
              controls (322.6139, 679.1894) and (340.7143, 569.5022) .. (343.8801, 500.7451)
              .. controls (346.1545, 451.3482) and (339.8855, 345.5133) .. (371.7373,
              357.5570) .. controls (419.5206, 374.5510) and (375.8183, 683.1693) ..
              (426.7141, 696.2945) .. controls (488.8102, 712.3080) and (381.4679, 389.6775)
              .. (515.2524, 431.2640);
            \path[fill=cffffff,line width=1px] (150.9074683464567, 519.595011023622)
              rectangle (193.99354582677168, 478.9026066141731);
            \path[fill=cffffff,line width=2.741px] (216.4338368503937, 552.2994897637794)
              rectangle (259.5199143307087, 478.7959218897637);
            \path[fill=cffffff,line width=2.741px] (264.68192503937007, 552.0381732283464)
              rectangle (307.768002519685, 478.53460535433067);
            \path[fill=cffffff,line width=2.256px] (322.5586507086614, 528.4465511811022)
              rectangle (365.64472818897633, 478.6391168503936);
            \path[fill=cffffff,line width=2.951px] (376.425282519685, 563.6257511811023)
              rectangle (419.51135999999997, 478.4433788976377);
            \path[fill=cffffff,line width=3.391px] (433.7331779527559, 587.937486614173)
              rectangle (477.91943433070867, 478.2126954330707);
          
            \path[fill=cffffff,line width=0.541px] (466.7967496062992, 488.8948157480314)
              rectangle (478.42907943307085, 478.26800050393695);
            \path[fill=black,line width=1px] (181.6796, 520.0327) circle (4px);
            \path[fill=black,line width=1px] (232.1928, 551.4877) circle (4px);
            \path[fill=black,line width=1px] (299.7561, 551.7814) circle (4px);
            \path[fill=black,line width=1px] (341.3936, 529.2701) circle (4px);
            \path[fill=black,line width=1px] (398.2643, 562.7832) circle (4px);
            \path[fill=black,line width=1px] (445.8939, 586.4382) circle (4px);
            \path[fill=black,line width=1px] (115.0955,383.5113) node[above right]
              (text2810){$A$};
            \path[fill=black,line width=1px] (517.2528,434.3439) node[above right]
              (text2814){$^UA$};
            \path[fill=black,line width=1px] (581.6259,511.8365) node[above right]
              (text2818){$\mathcal{M}(^U A) = f$};
          \end{tikzpicture}
          \caption{\light{The actual situation with the gauge fixing condition, where each orbit intersects multiple times, the number of intersections also being orbit dependent}}
    \end{subfigure}
    \caption{{The ideal and real gauge fixing situation}}
    \end{figure}

The presence of Gribov copies is closely tied to the existence of zero modes of the Faddeev-Popov operator. Mathematically, it means the existence of \(\alpha(x)\) (we are working back in Landau gauge) such that 

\begin{equation}
    \del_\mu (\del_\mu + f^{klm}A^l_\mu) \alpha(x) = 0
\end{equation}

This equation does have trivial solutions of the form \(\alpha(x) = c\mathbb{I}\) where \(\mathbb{I}\) is the identity element in the gauge group and \(c\) is a constant independent of spacetime. These solutions correspond to global gauge transformations and hence do not contribute to the problem of Gribov copies. The only relevant solutions are the ones that are non-trivial in the sense stated above. \\

From this, we can also see that in the case of QED, the above condition reduces to the Laplacian in \(\alpha\), \(\del^2\alpha = 0\), since \(f = 0\) for \(U(1)\) group. The solutions to this equation are the plane wave solutions, but these do not respect the boundary condition \(\alpha(\pm \infty)= c\) and hence these \(\alpha\) do not belong to the subset of allowed gauge transformations. Thus there exist no zero modes of the FP operator and hence QED is free of Gribov copies. \\
The same analysis also applies to the high momentum regime of the theory and also to the perturbative calculations, where the gauge fields are small perturbations around zero. In this case, again the FP operator approximately reduces to the Laplacian, and Gribov copies become irrelevant to the analysis. But in the low momentum regime, where large \(A\) effects become prominent, the theory is haunted by Gribov copies. 

\section{The Gribov Region}
% We note that the zero modes of the Faddeev-Popov operator imply that for that specific configuration, there exists an equivalent small gauge transformation for which both the original fields and the gauge transformed fields satisfy the gauge fixing condition. \\
If we consider the configuration space of the gauge fields, then in small perturbations around the vacuum, the FP operator is essentially the Laplacian, and hence it is positive definite. As we move away from the vacuum in any direction, at some distance there appears a zero mode. \\

Let us call this surface at which the Faddeev-Popov operator obtains its first zero mode the \textit{Gribov horizon}, and the region inside this horizon - the \textit{Gribov region}. Beyond this region, the FP operator is not ensured to be positive-definite and will have both negative modes and zero modes.\\

Inside the region, the FP operator has neither zero modes nor negative modes. If one restricts the path integral to this region alone, then both the problems encountered above would be resolved. Such a restriction would allow us to perform the gauge fixing and introduce the ghosts consistently.\\

One might ask what would be the consequences of such a restriction. Specifically one might ask if any non-trivial configurations are left out of the path integral by this restriction. This question was answered partially by Gribov himself, who showed that for every field infinitesimally close to the Gribov horizon, there exists a Gribov copy outside the horizon. 
The general statement of whether all orbits intersect the Gribov region or not is constructed using the Hilbert square norm along the orbit, and it is discussed below.\\

Thus the Gribov region can be defined as 
\begin{equation}
    C_0 = \{~A^k_\mu~|~ \del_\mu A^k_\mu = 0, \del_\mu D_\mu > 0~\}
\end{equation} 
and it has the following properties: 
\begin{enumerate}
    \item \textbf{Every orbit intersects the Gribov region at least once:}\\
    Consider a Hilber square norm on a given gauge orbit of the form 
    \begin{equation}
    ||A||^2 = \int dx~\text{Tr}(A_\mu A_\mu) = \frac{1}{2}\int dx ~A^k_\mu A^k_\mu 
    \end{equation}
    As we vary the configuration along the gauge orbit, we obtain different values for the Hilbert square norms. This norm attains its extremum along a given orbit when 
    \begin{equation}
        \begin{split}
            \delta ||A||^2 &= 2\int dx~\text{Tr}(\delta(A_\mu) A_\mu)= 0\\
            & = 2\int dx ~\text{Tr}((D_\mu\alpha)~A_\mu) = 0\\
            & = 2\int dx~ \text{Tr}(\del_\mu\alpha~ A_\mu  + A_\mu\alpha A_\mu - \alpha A_\mu A_\mu)=0\\
            & = 2\int dx~ \text{Tr}(\del_\mu\alpha~ A_\mu  + \alpha  A_\mu A_\mu - \alpha A_\mu A_\mu)=0\\
            & = \int dx~(\del_\mu \alpha^k) A_\mu^k = 0\\
            & = -\int dx~ \alpha^k (\del_\mu A_\mu^k) = 0
        \end{split}
    \end{equation}
    This implies that the configuration is an extremum only when \(\del_\mu A_\mu^k = 0\), i.e. only when the gauge fixing condition is satisfied. \\
    Further, the norm obtains a minima when 
    \begin{equation}
        \begin{split}
            \delta^2 ||A||^2 &= \delta\int  dx~ (-\alpha^k (\del_\mu A_\mu^k)) > 0\\
            &= \int dx~\alpha^k(-\del_\mu D^{kl}_\mu )\alpha^l > 0\\
        \end{split}
    \end{equation}
    This implies that the norm obtains a minimum when the FP operator is positive. Thus the Gribov region can be defined as the set of all minima of the Hilbert space norm.\\
    In~\cite{dellantonioEveryGaugeOrbit1991}, the authors have rigorously proved that the Hilbert space norm for each orbit will certainly attain a global minimum, hence proving that all gauge orbits intersect the Gribov region at least once. 
    \item \textbf{The trivial vacuum belongs to the Gribov region}: \\
    Since around \(A = 0\), the FP operator is essentially the Laplacian which is always positive definite, the vacuum and perturbations of the vacuum belong to the Gribov region. This means that all the perturbation theory calculations are free of the Gribov problem.
    \item \textbf{The Gribov region is convex}:\\
    What this means is that given two configurations \(A_1~\&~ A_2\) belonging to the Gribov region, then the affine combination \(A = \tau A_1 + (1-\tau)A_2\), where \(\tau\in [0,1]\), also belongs to the Gribov region.\\
    This can be easily seen as 
    \begin{equation}
        \del_\mu A_\mu = \tau\del_\mu A_{1\mu} + (1-\tau)\del_\mu A_{2\mu} = 0
    \end{equation}
    and 
    \begin{equation}
        \begin{split}
            \del_\mu D_\mu[A] &= \del_\mu(\del_\mu + f(\tau A_{1\mu} + (1-\tau)A_{2\mu}))\\
            & = \del_\mu(\tau \del_\mu + \tau fA_{1\mu} + (1-\tau)\del_\mu + (1-\tau) f A_{2\mu})\\
            & = \tau \del_\mu D_\mu[A_{1\mu}] + (1-\tau)\del_\mu D_\mu[A_{2\mu}]
        \end{split}
    \end{equation}
    Since both \(\tau\) and \(1-\tau\) are positive, \(-\del_\mu D_\mu [\tau A_{1\mu} + (1-\tau)A_{2\mu}]\) is positive. The above two conditions imply that if \(A_1\) and \(A_2\) belong to the Gribov region, then their affine combination also belongs to the Gribov region. 
    \item \textbf{The Gribov region is bounded:}\\
    Since both \(A\) and \(\alpha\) belong to the adjoint representation, \([A, \alpha]\) is traceless, and the sum of all eigenvalues of this for a certain \(A\) is zero. This implies that there exists at least one eigenvector \(\omega\) with eigenvalue \(\kappa<0\),such that \(\int dx dy~ \omega(x)[A(y), \omega(y)] = \kappa\)\\
    Now consider a field \(A\) residing inside the Gribov region. For a another field \(A'=\lambda A\), with positive \(\lambda\), 
    \begin{equation}
        \begin{split}
            \int dx dy ~ \omega(x) \del_\mu D_\mu[A] \omega(y) &= \int dx dy ~\omega(x)\del^2\omega(y) + \int dx dy~ \omega(x) [\lambda A(y), \omega(y)] \\
            &=\int dx dy ~\omega(x)\del^2\omega(y) + \lambda \kappa
        \end{split}
    \end{equation}
    since\([\lambda A, \alpha] = \lambda [A, \alpha]\), they both have the same eigenvectors but with eigenvalues scaled by \(\lambda\) \\
    For a sufficiently large \(\lambda\), the second term in the above expression becomes larger than the first term, and hence the overall sign of the inner product becomes negative, which means that the FP operator is not positive-definite. \\
    What this means is that for an \(A\) in the Gribov region and large enough \(\lambda\), \(A'=\lambda A\) live outside the Gribov region, which implies that the Gribov region is bounded on all sides. 
\end{enumerate}
With these properties, one might be tempted to conclude that restricting the path integral to the Gribov region not only resolves the problems with the introduction of the ghost fields but also removes Gribov copies from the integral. But there is one shortcoming to this argument. 
Since every minima of the Hilbert square norm on the gauge orbit lies in the Gribov region, there still is a possibility that some orbits have multiple local minima which will all enter the Gribov region, thus introducing copies in the region. 
This problem can be resolved if one considers another region, contained inside the Gribov region - the \textit{fundamental modular region}, containing only those configurations that are global minima of the Hilbert square norm. Since each gauge orbit has a unique global minima \cite{zwanzigerFundamentalModularRegion1994}, 
restriction to the fundamental modular region truly resolves the problem of Gribov copies. But so far nobody has managed to construct a path integral with the configuration space restricted to the fundamental modular region.\\

Nevertheless, the restriction of the configuration space to the Gribov region solves the non-positive-definiteness problem of the Faddeev-Popov operator. Thus such a restriction is not only possible but also necessary to define the path integral of the quantum Yang-Mills theory.  

\section{Gribov's No-Pole Condition} 
The inverse of the FP operator is nothing but the full propagator of the ghost fields. This means that the positive definiteness condition on the Faddeev-Popov operator translates to the condition that the ghost propagator should be positive and not have any non-trivial poles. \\

% (The following analysis in the specific case of \(\mathcal{G} = SU(N)\), thus  \(\dim(R) = N\) and \(\dim(\mathcal{G}) = N^2 - 1\).)\\

To look at this condition more closely, we consider the ghost propagator in Fourier space (without restriction to the Gribov region)
\begin{equation}
    G(p^2) = \int \mathcal{D}A ~\left\langle \bar\Omega(-p) \biggr | \frac{1}{-\del_\mu D_\mu}\biggr |\Omega(p)\right\rangle \det(\del_\mu D_\mu) \exp(-S[A])
    \label{eq:naiveGhostPropagator}
\end{equation}
Naive perturbation theory calculation of the above gives the result 
\begin{equation}
    G(p^2) = \frac{1}{p^2} \frac{1}{\left( 1 - \displaystyle\frac{11g^2 N}{48\pi^2} \ln \frac{\Lambda^2}{p^2}  \right)^{(3/22)(3/2 - \alpha/2)}}
\end{equation}
Where \(\Lambda\) is the ultraviolet cutoff and \(\alpha\) is the gauge parameter.\\

For large \(p^2\) the above expression is both non-zero and positive real.
But as we go to lower \(p\), problems start arising. \\
From the expression \ref{eq:naiveGhostPropagator}, poles in the propagator imply that we are approaching the surface where the FP operator has zero modes. 
The trivial pole \(p^2 = 0\) can be interpreted as approaching the Gribov horizon and is hence harmless. But the non-trivial pole \(\displaystyle p^2 = \Lambda^2 \exp\left(-\displaystyle \frac{48\pi^2}{11g^2N}\right)\) implies that we are approaching other surfaces with zero modes of Faddeev-Popov operator which means we are outside the Gribov region. 
Another noteworthy observation is that for values of \(\displaystyle p^2 < \Lambda^2 \exp\left(-\frac{48\pi^2}{11g^2N}\right)\), the propagator is non-real which also indicates that one has left the Gribov region. \\

Thus the restriction of the path integral to the Gribov region should imply that the non-trivial pole of the ghost propagator should be removed, and the propagator should be real and positive. We can do so by looking at the explicit expression of \(G(p^2)\), which we calculate by considering the gauge fields as an external coupling. We do this by summing the Feynman diagrams, i.e., we consider the following series

\begin{center}
    $G(p^2, A)^{kl}$ = \raisebox{-0.275ex}{\begin{tikzpicture}[baseline = (i1.base)]
        \begin{feynman}
            \vertex (i1) ;
        \vertex [right = of i1](a) {$\bar{\Omega^{k}}$};
        \vertex [right=of a] (b) {$\Omega^l$};
    
        \diagram* {
            (a) -- [ghost, momentum'=$p$, ] (b)
        };
        \end{feynman}
    \end{tikzpicture}} + \begin{tikzpicture}[baseline = (b.base)]
        \begin{feynman}
        \vertex (a) {$A^m$};
        \vertex [below= of a] (b);
        \vertex [left=of b] (i1) {$\bar{\Omega}^k$};
        \vertex [right=of b] (f1) {$\Omega^l$};
    
        \diagram* {
            (a) -- [gluon, momentum=$q-p$, ] (b),
            (i1) -- [ghost, momentum'=$p$] (b) -- [ghost, momentum' = $q$] (f1),
    
        };
        \end{feynman}
    \end{tikzpicture} + \begin{tikzpicture}[baseline = (b.base)]
        \begin{feynman}
        \vertex (a) {$A^m$};
        \vertex [below= of a] (b);
        \vertex [left=of b] (i1) {$\bar{\Omega}^k$};
        \vertex [right=of b] (c);
        \vertex [right=of a] (d) {$A^n$};
        \vertex [right=of c] (f1) {$\Omega^l$};

        \diagram* {
        (a) -- [gluon, rmomentum=$q$] (b),
            (d) -- [gluon, rmomentum=$q'$] (c),
            (i1) -- [ghost, momentum'=$p$] (b) -- [ghost, momentum'=$p-q$] (c) -- [ghost, momentum'=$p-q-q'$] (f1),
        };
        \end{feynman}
    \end{tikzpicture} 

\end{center}

and do term-by-term evaluation of it (here we consider terms only up to second order in \(g\) )\\

Since the external coupling should be integrated out, the gauge lines should be connected. Thus the second term can not exist and in the third term the two gauge ends are connected and \(q' = -q\). \\Thus the evaluation of the sum reduces to the evaluation of the following diagrams 
 
\begin{enumerate}
    \item \begin{tikzpicture}[baseline = (i1.base)]
        \begin{feynman}
            \vertex (i1) ;
            \vertex [right = of i1](a) {$\bar{\Omega^{k}}$};
            \vertex [right=of a] (b) {$\Omega^l$};
        
            \diagram* {
                (a) -- [ghost, momentum'=$p$, ] (b)
            };
        \end{feynman}
        \end{tikzpicture}\\
        This would be the free propagator for the ghost fields, which is simply 
        \begin{equation}
            G(p^2,A)~~(0^\text{th}~\text{order}) = \frac{\delta_{kl}}{p^2}
        \end{equation}
    \item \begin{tikzpicture}[baseline = (b.base)]
        \begin{feynman}
            \vertex (a);
            \vertex [below= of a] (b) ;
            \vertex [above left = 0.05em and 0.2em of b] (l1u) {$A^m_\mu$};
            \vertex [below= 0.1em of b] (l1l) {$\Omega^n$};
            \vertex [left=4em of b] (i1) {$\bar{\Omega}^k$};
            \vertex [right=6em of b] (c);
            \vertex [above right = 0.05em and 0.2em of c] (l2u) {$A^o_\mu$};
            \vertex [below= 0.1em of c] (l2l) {$\bar\Omega^p$};
            \vertex [right=of a] (d) ;
            \vertex [right=4em of c] (f1) {$\Omega^l$};

            \diagram* {
                (b) -- [gluon, momentum={[arrow shorten=0.25] $q$}, half left] (c),
                (i1) -- [ghost, momentum'={[arrow shorten=0.25] $p$}] (b) -- [ghost, momentum'={[arrow shorten=0.25]$p-q$}] (c) -- [ghost, momentum'={[arrow shorten=0.25] $p$}] (f1),
            };
            \end{feynman}
        \end{tikzpicture}\\
        This diagram can be evaluated as 
        \begin{equation}
            \begin{split}
                G(p^2,A)&~~(2^\text{nd}~\text{order})\\ 
                & = (-i)^2g^2 \frac{1}{V}\int \left(\frac{dq}{2\pi}\right)^4\frac{1}{p^2}\frac{\delta_{np}}{(p-q)^2}\frac{1}{p^2} f^{kmn}(p-q)_\mu A_\mu^m(-q) ~f^{pol} p_\nu A_\mu^o(q)\\
                & = -g^2\frac{1}{V}f^{kmn}f^{nol}\frac{1}{p^4} \int \left(\frac{dq}{2\pi}\right)^4\frac{1}{(p-q)^2} (p-q)_\mu p_\nu A^m_\mu(-q) A^o_\nu(q)
            \end{split}
        \end{equation}
        where the volume \(V\) has been introduced to maintain the dimensionality of the Feynman graph, since \(A\) has been introduced as external coupling.
    \end{enumerate}
To get the correct normalization, we trace over the color indices \(k~\&~l\) and divide by \(N^2 - 1\).\\
From this, we obtain the propagator (up to the second order in \(g\)) as 
\begin{equation}
    \begin{split}
        G(p^2,A) &= \frac{1}{p^2} -g^2\frac{1}{V} \frac{-N\delta_{mo}}{N^2-1}\frac{1}{p^4} \int  \left(\frac{dq}{2\pi}\right)^4\frac{1}{(p-q)^2} (p-q)_\mu p_\nu A^m_\mu(-q) A^o_\nu(q)\\
        &=  \frac{1}{p^2} + g^2 \frac{1}{V}\frac{N}{N^2-1}\frac{1}{p^4} \int \left(\frac{dq}{2\pi}\right)^4\frac{1}{(p-q)^2} (p-q)_\mu p_\nu A^m_\mu(-q) A^m_\nu(q)\\
        &= \frac{1}{p^2}\left( 1 + g^2\frac{1}{V}\frac{1}{p^2}  \frac{N}{N^2-1}\int \left(\frac{dq}{2\pi}\right)^4\frac{1}{(p-q)^2} (p-q)_\mu p_\nu A^m_\mu(-q) A^m_\nu(q)\right)
    \end{split}
\end{equation}
Calling \(\displaystyle \sigma(p^2,A) = g^2\frac{1}{V}\frac{1}{p^2}  \frac{N}{N^2-1}\int \left(\frac{dq}{2\pi}\right)^4\frac{1}{(p-q)^2} (p-q)_\mu p_\nu A^m_\mu(-q) A^m_\nu(q) \), the propagator, up to second order in \(g\), becomes
\begin{equation}
    G(p^2,A) = \frac{1}{p^2}(1+\sigma(p^2,A))
\end{equation}
Because we are summing over the entire series of Feynman diagrams, the complete propagator can be appropriately approximated as 
\begin{equation}
    G(p^2,A) \approx \frac{1}{p^2}\frac{1}{1-\sigma(p^2, A)}
\end{equation}
From this, the no-pole condition simply reads as \(\sigma(p^2,A)<1\).\\

This condition can be simplified further by considering that since we are working in the Landau gauge, the transversality condition should hold, i.e. \(q_\mu A_\mu(q) = 0 \).\\
With this gauge, we can write 
\begin{equation}
    A_\mu(-q)A_\nu(q) = C(A) \left( \delta_{\mu\nu} - \frac{q_\mu q_\nu}{q^2}\right)
\end{equation}
To find the form of \(C(A)\), we multiply the above equation by \(\delta_{\mu\nu}\) and see that \(A_\mu(-q)A_\mu(q) = C(A)(4-1)\). Therefore
\begin{equation}
    A_\mu(-q)A_\nu(q) = \frac{1}{3} A_\rho(-q)A_\rho (q) \left( \delta_{\mu\nu} - \frac{q_\mu q_\nu}{q^2}\right)
\end{equation}
With this, and using the transversality condition,
\begin{equation}
    \sigma(p^2,A) = g^2 \frac{1}{V}\frac{N}{N^2-1}\frac{1}{3} \frac{p_\mu p_\nu}{p^2}\int \left(\frac{dq}{2\pi}\right)^4\frac{1}{(p-q)^2}A_\rho(-q)A_\rho (q) \left( \delta_{\mu\nu} - \frac{q_\mu q_\nu}{q^2}\right)
\end{equation}
Using \(\displaystyle \int d^4q~ f(q^2) \frac{q_\mu q_\nu}{q^2} = \frac{1}{4}\delta_{\mu\nu}\int d^4q f(q^2)\), \(\sigma\) can be written as 
\begin{equation}
    \sigma(p^2,A) = g^2 \frac{1}{V}\frac{N}{N^2-1}\frac{1}{4} \frac{p_\mu p_\nu}{p^2} \delta_{\mu\nu}\int \left(\frac{dq}{2\pi}\right)^4\frac{1}{(p-q)^2}A_\rho(-q)A_\rho (q)
\end{equation}
This is a continuously decreasing function of \(p^2\). Therefore the maximum value of \(\sigma\) is at \(p^2 = 0\). Thus, the no-pole condition can be restated as \(\sigma(0,A) < 1\), i.e.
\begin{equation}
    \begin{split}
        \sigma(0,A) &= \lim_{p^2\rightarrow 0} g^2 \frac{1}{V}\frac{N}{N^2-1}\frac{1}{4} \int \left(\frac{dq}{2\pi}\right)^4\frac{1}{(p-q)^2}A_\rho(-q)A_\rho (q) \\
        & =  g^2 \frac{1}{V}\frac{N}{N^2-1}\frac{1}{4} \int \left(\frac{dq}{2\pi}\right)^4\frac{1}{q^2}A_\rho(-q)A_\rho (q) < 1
    \end{split}
\end{equation}
Since this expression is independent of any momenta, the no-pole condition can be inserted into the path integral to restrict it to the Gribov region, in the form of Heaviside step function \(\theta(1-\sigma(A))\).

\section{Restricting the Path Integral Using No-Pole Condition}
It is pretty straightforward to implement the no-pole step function into the path integral, by using the integral representation of the theta function in the form 

\begin{equation}
    \theta(1-\sigma(A)) = \int_{-\infty + i\epsilon}^{\infty + i\epsilon} d\beta \frac{1}{2\pi i \beta}~\exp(\beta (1-\sigma(A)))
\end{equation}

Therefore the modified partition function would be 
\begin{equation}
    Z = \int \mathcal{D}A~ \int d\beta \frac{e^\beta}{2\pi i \beta}~ \exp\left(-S[A] - \beta\sigma[A] - J\cdot A\right)~\det(\del_\mu D_\mu)
\end{equation}
Since this additional term is quadratic in the gauge fields, we expect it to modify the form of the gauge propagator. To calculate the free gauge propagator, we consider only the gauge part of the path integral, quadratic in fields. \\
It is convenient to work in Fourier space where the action is\\
\begin{equation*}
    S_{quad}[A] = \int \frac{d^4q}{(2\pi)^4} A_\nu(-q) (~\delta_{\mu\nu}q^2 + q_\mu q_\nu\left(\frac{1}{\alpha}-1\right)~)A_\mu (q)  
\end{equation*}
and thus the partition function becomes
\begin{equation}
    \begin{split}
    Z_{A}[J] = \int d\beta \frac{e^\beta}{2\pi i \beta}~\int \mathcal{D}A~  \exp(-\int dq~A_\nu(-q) (~\delta_{\mu\nu}\left(q^2 + \beta g^2 \frac{1}{V}\frac{N}{N^2-1}\frac{1}{4} \frac{1}{q^2} \right ) \\
    +\left(\frac{1}{\alpha}-1\right) q_\mu q_\nu~)A_\mu (q) - J_\mu(-q) A_\mu(q))
    \end{split}
\end{equation}

Calling \(\displaystyle \mathcal{M}_{\mu\nu}^{kl} = \delta^{kl}\left(\delta_{\mu\nu}\left(q^2 + \beta g^2 \frac{1}{V}\frac{N}{N^2-1}\frac{1}{4} \frac{1}{q^2}\right) +\left(\frac{1}{\alpha}-1\right) q_\mu q_\nu \right)\), the gluon propagator can be easily calculated as

\begin{equation}
    \braket{A^k_\mu(-p)A^l_\nu(p)} = \int \frac{d\beta}{2\pi i\beta} e^\beta \frac{1}{\sqrt{\det(\mathcal{M})}}\mathcal{M}^{-1}
\end{equation}

Using \(\det\mathcal{M}\ = \exp\text{Tr}\ln\mathcal{M}\), we can write 
\begin{equation}
    \braket{A^k_\mu(-p)A^l_\nu(p)} = \int d\beta~ \exp\left(\beta - \ln \beta - \frac{1}{2}\text{Tr}\ln\mathcal{M} \right)\mathcal{M}^{-1}
\end{equation}
where all the constants have been absorbed into the normalization.\\

To perform the integration over \(\beta\), we use the method of steepest descent. Since the exponential is a monotonically increasing function, the most dominant term in the integral would be the one for which the exponential would be the maximum (assuming that \(\mathcal{M}^{-1}\)) doesn't oscillate much. The maximum value of \(\beta\) can be found as the solution to the gap equation 

\begin{equation}
    \frac{d}{d\beta}\left( \beta - \ln \beta - \frac{1}{2}\text{Tr}\ln\mathcal{M}\right)  = 0
\end{equation}

The integration over \(\beta\) can now be approximated as a simple substitution \(\beta \rightarrow \beta_0\) in the integrand. Thus, the modified propagator becomes 
\begin{equation}
    \braket{A^k_\mu(-p)A^l_\nu(p)} = \mathcal{M}^{-1}\biggr | _{\beta\rightarrow \beta_0}
\end{equation}
Where the exponential is dropped since it gets canceled by the normalization.\\

% With such a substitution, \(\mathcal{M}\) becomes 
% \begin{equation}
%     \delta_{kl} \left(~\delta_{\mu\nu}\left(q^2 + \beta_0 g^2 \frac{1}{V}\frac{N}{N^2-1}\frac{1}{4} \frac{1}{q^2} \right ) +\left(\frac{1}{\alpha}-1\right) q_\mu q_\nu~ \right)
% \end{equation}
Calling \(\displaystyle \beta_0 g^2 \frac{1}{V}\frac{N}{N^2-1}\frac{1}{4} \equiv \kappa^4\), and taking the limit \(\alpha \rightarrow 0\), the propagator in Landau gauge can be explicitly written as 
\begin{equation}
    \braket{A^k_\mu(-p)A^l_\nu(p)}  = \delta_{kl} \left( \frac{p^2}{p^4 + \kappa^4} \left( \delta_{\mu\nu} - \frac{p_\mu p_\nu}{p^2} \right)  \right)
\end{equation}
This is a propagator of a massive field, with a mass gap defined by \(\kappa\)
Thus the restriction to the Gribov region introduces a dynamic mass gap in the Yang-Mills theory. \\
This propagator has complex poles, \(p^2 = \pm i\kappa^2\), which leads to the violation of the reflection positivity axiom of the Osterwalder-Schrader axioms for quantum field theory. The violation of the positivity of the Schwinger function implies that the gluon states cannot enter the scattering matrix as asymptotic states. Such a theory is said to be confined.\\

The physical intuition behind the occurrence of the mass gap is pretty simple. Just like in an infinite space, the momentum spectrum is continuous, but once one introduces periodic boundary conditions on the said space, there arises a discretization of the momentum spectrum; the confinement of the functional integral to the Gribov region introduces a mass gap into the Yang-Mills theory.\\
Thus, by introducing the no-pole condition, we have not only introduced a mass gap into the Yang-Mills theory but also have made a way for confinement.

%%%%%%%%%%%%%%%%%%%%%%%%%
% Created on 03-10-2023 %
%                       %
% Author: Adithya A Rao %
%%%%%%%%%%%%%%%%%%%%%%%%%

\newpage

\part{Stochastic Quantization}

\chapter{Prelude - Probability Theory and Stochastic Processes}
\vspace{-45pt}
In this chapter, we discuss the fundamentals of probability theory and stochastic processes. Stochastic calculus deals with the study of stochastic processes involving randomness. It provides a rigorous mathematical framework for working with random variables and stochastic differential equations. We further discuss in-depth one of the most fundamental stochastic processes - the Brownian motion which models the erratic, unpredictable movement of a particle subjected to random perturbations.

\section{Stochastic Processes}
Consider a set \(\Omega\) whose elements are to be interpreted as the possible outcomes of a probabilistic experiment. On this set, we define a \(\sigma \)-algebra \(\mathcal{S}\), which is a set whose elements are \textit{some subsets} of \(\Omega \), such that the following properties are satisfied
\begin{itemize}
    \item \(\Omega \in \mathcal{S}\)
    \item \(\forall E \in  \mathcal{S}, \exists E^C \equiv \Omega \backslash E \in \mathcal{S}\). That is, for every element \(E\) of \(\mathcal{S}\), \(E^C\) which is defined as the complement of \(E\) in \(\Omega\) is also in the set \(\mathcal{S}\) 
    \item \(\forall E_n \in \mathcal{S}, n=1,2,\dots,~ \cup_n E_n \in \mathcal{S}\). That is, the finite union of elements in \(\mathcal{S}\) also belongs to \(\mathcal{S}\)
\end{itemize}
With such an algebra defined, we call the tuple \((\Omega, \mathcal{S})\) a measurable space.\\

A measure is a function defined on a measurable space. It is a function \(\mu:\mathcal{S}\to [0,~\infty)\), that satisfies 
\begin{itemize}
    \item If \(E_1 \subset E_2\), then \(\mu(E_1) < \mu(E_2)\).
    \item If \(E_n \in \mathcal{S},~n=1,2,\dots\) and \(E_i \cap E_j = \varnothing~(i\neq j)\), then \(\mu(\cup_n E_n) = \sum_n\mu(E_n)\)
\end{itemize}
If we further impose a condition \(\mu(\Omega) = 1\), then the measure so defined is called the probability measure and is usually denoted by \(P\). The tuple \((\Omega, \mathcal{S}, P)\) is then called the probability space.\\

Suppose \((\Omega, \mathcal{S})\) and \((\Omega', \mathcal{S'})\) are two measurable spaces. The function \(X:\Omega \to \Omega'\) is called a random variable, if for every \(A\in \mathcal{S'}\), \(X^{-1}(A)\in \mathcal{S}\). If the first measurable space is equipped with a probability measure \(P\), then the random variable induces a probability measure on the second space given by \(\mathcal{P} \equiv P\circ X^{-1}\), called the distribution (law) of \(X\).\\

In general, the second measurable space is taken to be \((\mathbb{R},~\mathfrak{B})\), where \(\mathfrak{B}\) is the Borel \(\sigma\)-algebra on \(\mathbb{R}\). For this case, the distribution of \(X\) is completely determined by the \textit{distribution function } defined as 
\begin{equation}
    F_X(x) = P(X\le x)
\end{equation}
where \(X\le x\) is defined as the set \(\{ X\le x \} = \{ \omega \in \Omega ~|~X(\omega)\le x  \}\).\\
The distribution function is continuously non-decreasing and gives the measure associated with \((x,~x+dx) \in \mathfrak{B}\) as \(d(F_X(x))\)\\
A special case is when the distribution function can be written in the form 
\begin{equation}
    F(x) = \int_{-\infty}^{x} p(x)dx
\end{equation}
In this case, the function \(p(x)\) is called the probability density function of \(X\), and the measure associated with \(dx\) is \(d(F_X(x) = p(x) dx\). \\

With the knowledge of the distribution function, one can define the expectation value of any function \(h(X)\) as 
\begin{equation}
    \mathbb{E}h(X) = \int h(x)d(F_X(x))
\end{equation}

With these definitions, we now define a stochastic process. A stochastic process is an indexed set of random variables \({X_t,~ t\in T}\), where \(T\) is called the index set and is usually taken (in the continuous case) to be \(T = [0, \infty)\). In a physical setting, the index is time and a stochastic process can be interpreted as, for each \(t\in T\), \(X_t\) picks an event \(E \in \mathcal{S}\) with probability \(P(E)\), and returns \(X_t(E)\). \\

To proceed, let us consider first a simple physical process that can be modeled by a stochastic process, and extend the model to quantize a given system too.

\section{Brownian Motion}
The Brownian motion is one of the oldest and simplest stochastic processes known to physicists. Physically it is a diffusion process of a particle in a medium under a static drift force \(K(x)\) while also accounting for the random collisions of the particle with the medium. \\

The forces on this diffusing particle are the external drift force, \(K(x)\), a friction force opposing the motion due to the mismatch in the number of collisions in the direction of motion and the opposite direction, \(-\alpha \dd{x}{t}\), and a gaussian noise force due to the random collisions the particle will have with the surrounding media, \(\alpha\eta(t)\).\\
The \(\eta(t)\)\hspace{1pt}s are random variables with Gaussian distributions, satisfying 
\begin{equation}
    \begin{split}
        \langle \eta(t) \rangle_\eta =& ~0\\
        \langle \eta(t')\eta(t'')\rangle_\eta = & ~2D\delta(t'-t'')
    \end{split}
\end{equation}
where \(\langle \cdot \rangle\) denores the expectation value with \(\displaystyle p(\eta) = \frac{1}{\mathcal{N}}\exp\left( -\frac{1}{4D}\int dt ~\eta(t)^2 \right)\).\\

The Newton's equation for this particle would be 
\begin{equation}
    m\frac{d^2x}{dt^2} = -\alpha \dd{x}{t} + K(x) + \alpha\eta(t)
\end{equation}
We consider the situation where the mass of the particle is negligible compared to the drift and friction forces. In this case, we can neglect the inertial term to give 
\begin{equation}
    \dd{x}{t} = \frac{1}{\alpha} K(x) + \eta(t)
    \label{eq:LangevinDiffusion}
\end{equation}
This is the Langevin equation.\\
Calling \(\dfrac{1}{\alpha}\) as \(\gamma\), for a given initial condition \(x(0) = x_0\), the above differential equation has the formal solution 
\begin{equation}
    x_{[\eta]}(t) = x_0 + \int_0^t \left( \gamma K(x) + \eta(t') \right)\text{dt}'
\end{equation}
This is a functional of \(\eta\), and since \(\eta\) is a stochastic process, \(x_{[\eta]}\) is also a stochastic process with \(x_{[\eta]}(t)\)s being random variables. We can define the  probability distribution for the position \(x\) at time \(t\) as 
\begin{equation}
    p(x,t) = \langle~  \delta(x - x_{[\eta]}(t))~ \rangle_\eta
\end{equation} 

This probability distribution follows the evolution equation 
\begin{equation}
    \pp{{p}(x, t)}{t} = \dd{}{x} \left( \left[ D\dd{}{x}- \gamma K(x) \right] {p}(x,t)\right)  
    \label{eq:LangevinProbDens}  
\end{equation}
called the Focker-Planck equation. \\

\subsection{Focker-Planck equation from the Langevin equation}

To see how the Focker-Planck equation follows from the Langevin equation, start by discretizing the Langevin equation
\begin{equation}
    x(t+dt) - x(t) = \gamma K(x)~dt + \frac{1}{\sqrt{dt}} \tilde{\eta}(t) dt
\end{equation}
where \(\langle \tilde{\eta} \rangle = 0\) and \(\langle \tilde{\eta_{t'}} \tilde{\eta_{t''}} \rangle = 2D\delta_{t't''}\)\\

(The \(\dfrac{1}{\sqrt{dt}}\) is needed since \(\delta(t'-t'')\) has dimension \(t^{-1}\), but \(\delta_{t't''}\) is dimensionless making \(\dfrac{1}{\sqrt{dt}}\tilde\eta\) the right gaussian random variable for the discrete version).\\

Now consider a generic functional of \(x(t)\), \(f(x(t))\). To compute \(\dd{f}{t}\), we obtain (upto first order in \(dt\))
\begin{equation}
    \begin{split}
        f(x(t+dt)) - f(x(t)) &= f(x(t) + \gamma K(x(t))dt + \sqrt{dt}\tilde\eta) - f(x(t))\\
        &=\gamma\pp{f}{x}Kdt + \pp{f}{x}\tilde\eta \sqrt{dt} + \frac{1}{2}\frac{\partial^2 f}{\partial x^2}\tilde\eta^2 dt
    \end{split}    
\end{equation}
Dividing by \(dt\) and taking the average over \(\tilde\eta\), we find that the second term becomes zero due to the zero mean of \(\tilde\eta\) and the equation becomes
\begin{equation}
    \left\langle \dd{f}{t} \right\rangle = \left \langle \pp{f}{x}\gamma K +   \frac{\partial^2f}{\partial x^2}D  \right \rangle
\end{equation}

Using \(\displaystyle \langle f(x) \rangle = \int f(x) p(x,t)  dx \), where \(p(x,t)\) is the probability distribution function in \(x\), we get the left hand side as
\begin{equation}
    \begin{split}
        \left \langle \dd{f}{t} \right \rangle &= \pp{}{t} \int f(x) p(x,t) dx = \int f(x) \pp{p(x,t)}{t}dx
    \end{split}
\end{equation}
and the right-hand side as 
\begin{equation}
    \begin{split}
        \left \langle \pp{f}{x}\gamma K +  \frac{\partial^2f}{\partial x^2}D  \right \rangle &= \gamma K(x) p(x,t) f(x) \Bigg|_{-\infty}^{\infty} - \int \gamma f(x) \pp{K(x)p(x,t)}{x} dx \\
        &~~~~~+ DP(x,t)\pp{f}{x}\Bigg|_{-\infty}^{\infty}- \int D \pp{f}{x}\pp{p(x,t)}{x}dx
    \end{split}
\end{equation}
Using the fact that \(p(x,t)\) goes to zero at both infinities, we get
\begin{equation}
    \left \langle \pp{f}{x}\gamma K +  \frac{\partial^2f}{\partial x^2}D  \right \rangle =- \int \gamma f(x) \pp{K(x)p(x,t)}{x} dx + \int D f(x) \frac{\partial^2 p(x,t)}{\partial x^2} 
\end{equation}

Since the above relation should hold for any function \(f(x)\), we see that it must hold that 
\begin{equation}
    \begin{split}
        &\int f(x) \left ( \pp{p(x,t)}{t} +  \gamma \pp{K(x)p(x,t)}{x} - D \frac{\partial^2 p(x,t)}{\partial x^2}  \right) dx = 0\\
        &\implies  \pp{p(x,t)}{t} = -  \gamma \pp{K(x)p(x,t)}{x} + D \frac{\partial^2 p(x,t)}{\partial x^2} \\
        & \implies \pp{p(x,t)}{t} = \pp{}{x}\left (  \left[ D\pp{}{x} - \gamma K(x) \right]p(x,t)\right)
    \end{split}
\end{equation}
which is the Focker-Planck equation.
\\

\subsection{The thermal equilibrium}

One important case of Brownian motion is when the drift force is conservative. That is, when \( K(x)\) can be written in terms of the gradient of a potential
\begin{equation}
    K(x) = - \pp{V(x)}{x}
\end{equation} 

Let us, for the moment, consider \(\gamma = 1\) and \(D = 1\). \\
The Focker-Planck equation then becomes 
\begin{equation}
    \pp{p}{t} = \pp{}{x}\left(  \pp{}{x} + \pp{V}{x}  \right)p
\end{equation}
Consider the distribution \(\displaystyle \psi(x,t) = p(x,t) \exp\left( \frac{V(x)}{2} \right)\) \\\(\displaystyle \implies p(x,t) = \psi(x,t)\exp\left( -\frac{V(x)}{2}  \right)\). \\
The distribution \(\psi\) therefore follows the equation 
\begin{equation}
    \begin{split}
        \exp\left( -\frac{V}{2}  \right)\pp{\psi}{t} &= \pp{}{x}\left( \exp\left( -\frac{V}{2}  \right)\pp{\psi}{x} + \exp\left( -\frac{V}{2}  \right)\frac{1}{2}\psi \pp{V}{x} \right)\\
        &= -\exp\left( -\frac{V}{2}  \right)\frac{1}{2}\pp{V}{x}\left( \pp{\psi}{x} + \frac{1}{2} \psi \pp{V}{x}  \right) \\
        &~~~~~~~~~~~~~~~~~~~~~~~~~+ \exp\left( -\frac{V}{2}  \right)\left( \pp{^2\psi}{x^2} + \frac{1}{2} \pp{\psi}{x}\pp{V}{x} + \frac{1}{2}\psi \pp{^2 V}{x^2} \right)\\
        \implies
        \pp{\psi}{t} &= \left( \pp{^2}{x^2} - \frac{1}{4} \left(\pp{V}{x} \right)^2  + \frac{1}{2} \pp{^2V}{x^2}  \right)\psi
    \end{split}
    \label{eq:FP_for_psi}
\end{equation}
Calling
\begin{equation}
    H = -\frac{1}{2} \left( \pp{^2}{x^2} - \frac{1}{4} \left(\pp{V}{x} \right)^2  + \frac{1}{2} \pp{^2V}{x^2}  \right) 
\end{equation}
we get 
\begin{equation}
    \pp{\psi}{t} = -2H\psi
    \label{eq:SchrodingerType}
\end{equation}
which is a Schrodinger-type equation.\\

Since we can write 
\begin{equation}
    H = \frac{1}{2}\left( -\pp{}{x} + \frac{1}{2}\pp{V}{x}   \right)\left( \pp{}{x} + \frac{1}{2}\pp{V}{x}   \right)
\end{equation}
Since \(\pp{}{x}^T = -\pp{}{x}\), and \(\pp{V}{x}\) is proportional to the identity operator, we see that the operator \(H\) is self-adjoint, and is in the form \(H = M^T M\). \\

Therefore the operator \(H\) has a non-negative real spectrum, with the eigenvectors forming a complete orthonormal basis.
Therefore, we can write the solution to the equation \ref{eq:SchrodingerType} as 
\begin{equation}
    \psi(x,t) = \sum_{n=0}^\infty a_n \psi_n(x) \exp\left( -E_n t \right) = \psi_0(x) + \sum_{n=1}^\infty a_n\psi_n(x) \exp(-E_n t)
\end{equation}
where \(\psi_n(x)\) form an orthonormal basis with \(H\psi_n(x) = E_n \psi_n(x)\) and \(\psi_0\) is the eigenfunction with \(E_0 = 0\).\\

We immediately see that \(\psi_0\) follows (Since \(A^T Ax =0 \implies Ax = 0\)),
\begin{equation}
    H\psi_0(x) = 0 \implies \left(  \pp{}{x}  + \frac{1}{2}\pp{V}{x} \right)\psi_0 = 0 \implies \psi_0 \propto \exp\left( -\frac{V}{2} \right)  
\end{equation}

Therefore the original probability distrubution \(p\) has the solution 
\begin{equation}
    p(x,t) = \psi \exp\left( -\frac{V}{2} \right) = a_0\exp\left( -V \right) + \sum_{n=1}^\infty a_n \psi_n(x)\exp\left( -\frac{V}{2} \right)\exp(-E_n t)
\end{equation}

At thermal equilibrium, i.e. in the limit \(t\to \infty\), all terms with \(n>0\) become zero, while only the \(n=0\) term survives, giving the equilibrium probability distribution as 
\begin{equation}
    p_{\text{eq}}(x) = \lim_{t\to \infty} p(x,t) = a_0\exp(-V(x))
\end{equation}
where the constant \(a_0\) is fixed by normalisation, i.e. \(a_0^{-1} = \displaystyle\int_{-\infty}^{\infty} \exp(-V(x))dx\)\\
The similarity of the structure of this probability distribution to the path integral of quantum field theory is what allows one to define a stochastic quantization prescription that allows one to view quantum mechanics as a Brownian motion but in configuration space.  
%%%%%%%%%%%%%%%%%%%%%%%%%
% Created on 03-10-2023 %
%                       %
% Author: Adithya A Rao %
%%%%%%%%%%%%%%%%%%%%%%%%%

\newpage
\chapter{Quantum Field Theory as a Brownian Motion}
\vspace{-45pt}
In this chapter, we show how a quantum field theory can be written as an equilibrium limit of a Brownian motion in the configuration space. Developed in the early 1980s by Gert Parisi and Yong-Shi Wu, unlike traditional quantization methods, which rely on the path integral formulation and the introduction of ghost fields, stochastic quantization reformulates the quantum field theory as a stochastic process described by a Langevin equation. We explicitly quantize the scalar field theory with stochastic quantization prescription, and also discuss the stochastic quantization for gauge fields and the problems associated with this formalism.

\section{Stochastic Quantization}
There is a striking similarity in the equilibrium distribution of the position of a particle undergoing Brownian motion and the Euclidean Quantum Field Theory measure, the correspondence being 
\begin{equation}
    Z_0^{-1}\exp\left(-\frac{S(\phi)}{\hbar}\right) \iff a_0~\exp\left(-\frac{V(x)}D\right) ~~\text{with}~~D \iff \hbar
\end{equation}
This motivates a generalization of the Brownian motion to an infinite dimensional configuration space, expecting the equilibrium limit of the probability distribution to be the same as the Euclidean Quantum Field Theory measure.\\
Note that the probability distribution obtained in the case of Brownian motion was at time \(t\to \infty\), but in Euclidean Quantum Field Theory the distribution is expected for all \(t\) and not only at \(t\to \infty\). Therefore we see that there is a lack of a \textit{physical} parameter that can be sent to infinity to obtain the distribution. To overcome this, we introduce a fictitious \textit{time} parameter, in which a Brownian motion is supposed, and the equilibrium limit is taken to obtain the Euclidean Quantum Field Theory measure. \\

Therefore, formally, the stochastic quantization prescription is to take a classical field theory with fields \(\phi(x)\) (the \(x\) here stands now for both space and physical time dimensions) and action \(S[\phi]\) and extend the definition of the fields as \(\phi(x)\to \phi_\eta(x,t)\), where \(t\) is now the fictitious time parameter. With the action as a drift potential, construct a Langevin equation for the fields as 
\begin{equation}
    \pp{\phi_\eta(x,t)}{t} = -\DD{S[\phi_\eta]}{\phi_\eta(x,t)} + \eta(x,t)
\end{equation}
with initial condition \(\phi_\eta(x,0) = C(x)\) where \(C(x)\) is a function independent of \(t\).\\
The correlation functions of Euclidean Quantum Field Theory become averages on \(\eta\), 
\begin{equation}
    \braket{\phi_\eta(x_1, t_1)\dots\phi_\eta(x_n, t_n)}_\eta = \int \mathcal{D}\eta ~\exp\left( -\frac{1}{4}\int \eta^2(x,t)~dxdt  \right)~ \phi_\eta(x_1, t_1)\dots\phi_\eta(x_n, t_n)
\end{equation}
which, in the limit \(t\to \infty\), gives the desired Euclidean Quantum Field Theory correlation functions as 
\begin{equation}
    \lim_{t\to \infty} \braket{\phi_\eta(x_1, t_1)\dots\phi_\eta(x_n, t_n)}_\eta = \braket{\phi_\eta(x_1)\dots\phi_\eta(x_n)}
\end{equation}

For this, we can also write the corresponding Focker-Planck equation as 
\begin{equation}
    \begin{split}
        &\pp{P(\phi,t)}{t} = \int dx~\DD{}{\phi(x,t)} \left(  \DD{}{\phi(x,t)} + \DD{S[\phi]}{\phi(x,t)}\right) P(\phi, t)\\
    \end{split}
\end{equation}
with initial condition \(P(\phi, 0) = \prod_y \delta(\phi(y))\). \\

The equilibrium limit \(t\to \infty\) of \(P(\phi, t)\) is expected to give the required Euclidean Quantum Field Theory distribution,
\begin{equation}
    \underset{t\to \infty}{\mathrm{w.lim}} ~P(\phi, t) = \frac{\mathrm e^{-S}}{\int \mathcal{D}\phi~ \mathrm{e}^{-S}}
\end{equation}
where the \(\mathrm{w.lim}\) denotes a weak limit which means that the limit is taken only when the probability density is applied to a string of fields. \\

In the following section, we show explicitly how the scalar field theory can be quantized via stochastic quantization prescription.

\section{Quantizing the Scalar Fields}
For a scalar field theory governed by the action \(S = \displaystyle \int d^4 x~  \left[\frac12 (\del\phi(x))^2 - \frac12(m\phi(x))^2  \right] \) the Langevin equation is 
\begin{equation}
    \pp{\phi_\eta}{t}  = (\del^2  - m^2)\phi_\eta + \eta(x, t)
\end{equation}

Consider the Fourier Transformed Langevin equation in \(k\) and \(t\) given as 
\begin{equation}
    \pp{}{t}\phi_\eta(k,t)  + (k^2  + m^2)\phi_\eta(k,t) = \eta(k, t)
\end{equation}
One can obtain \(\phi_\eta\) by using Duhamel's formula, 
\begin{equation}
    \phi_\eta(k,t) = \int_0^t G(k,~t')~\eta(k, t')~dt'
    \label{eq:Duhamel}
\end{equation}

where \(G(k,~ t-t')\) is the solution to the Fourier transformed heat equation 
\begin{equation}
    \pp{}{t}G(k,t)  + (k^2  + m^2)G(k,t) = 0
\end{equation}
which is given by 
\begin{equation}
    G(k,t) = \mathrm{e}^{-(k^2 + m^2)(t-t')}
\end{equation}
Therefore, using this in eq \ref{eq:Duhamel} and taking inverse Fourier transform, we get 
\begin{equation}
    \phi_\eta(x,t) = \frac{1}{(2\pi)^d}\int_0^t\int~\mathrm{e}^{ikx} ~\mathrm{e}^{-(k^2 + m^2)(t-t')}~\eta(k, t')~dk~dt'
\end{equation}
which from the convolution property of the Fourier transform becomes 
\begin{equation}
    \phi_\eta(x,t) =\int  \int_0^t \left(\frac{1}{(2\pi)^d} \int ~\mathrm{e}^{ik(x-y)-(k^2 + m^2)(t-t')} dk~ \right) \eta(y, t)~dt'~dy
\end{equation}

where the function 
\begin{equation}
    G(x-y, ~t-t') =\frac{1}{(2\pi)^d} \int ~\mathrm{e}^{ik(x-y)-(k^2 + m^2)(t-t')} dk
\end{equation}
can be seen as the Green function for the free Langevin equation.

The propagator can be calculated as a two-point correlator as 
\begin{equation}
    D(x,y;t,t') = \langle \phi_\eta(x,t)\phi_\eta(y,t')\rangle_\eta
\end{equation}
which can be evaluated as 
\begin{equation}
    \begin{split}
        D &=  \int dz \int dz' \int_0^t d\tau \int_0^{t'} d\tau' G(x-z, t-\tau)G(y-z', t'-\tau')\left  \langle \eta(z,\tau)\eta(z',\tau') \right \rangle_\eta\\
        &= 2\int dz \int dz' \int_0^t d\tau \int_0^{t'} d\tau' G(x-z, t-\tau)G(y-z', t'-\tau') ~\delta(z-z')\delta(\tau-\tau')
    \end{split}
\end{equation}
The integration over the time parameter should be done first on the one whose integration range is larger since doing it the other way will leave parts of the range where the delta function's argument is not zero. Therefore, the above integration becomes 

\begin{equation}
    D =  2\int dz\int_0^{\min(t,t') }d\tau~ G(x-z, t-\tau)G(y-z, t'-\tau)
\end{equation}

Let us, for a moment, consider the probability density obtained from this theory. Writing \(P_0(\phi, t)\) as 
\begin{equation}
    P_0[\phi, t] = \int \mathcal D\eta \int \mathcal{D}k~\exp\left( -\frac{1}{4}\int \eta^2(x',t') ~dx' dt' \right)  \exp\left(i\int dx ~k(x)~(\phi(x) - \phi_\eta(x, t))\right)
\end{equation}
which we can evaluate now, using the form of \(\phi_\eta\) as
\begin{equation}
    \begin{split}
        P_0[\phi, t] =& \int \mathcal D\eta \int \mathcal{D}k~\exp\left( -\frac{1}{4}\int \eta^2(x,t) ~dx dt \right.\\
        &~~~~~~~~~~~~~~~~~~~~\left. +i\int dx ~k(x)~\left(\phi(x) -\int  \int_0^t G(x-y,~t-t') \eta(y, t)~dt'~dy\right)\right)        
    \end{split}
\end{equation}  

Performing the Gaussian integral over \(\eta\), we obtain 
\begin{equation}
    \begin{split}
        P_0[\phi, t] =& \mathcal{N}^{-1} \int\mathcal{D}k~\\
        &~~~~~~~~\exp\left( -\int dx_1 dx_2~ k(x_1) \int dy_1 dy_2 \int_0^{t} d\tau_1 d\tau_2~ G(x_1-y_1, t -\tau_1)\right.\\
        &~~~~~~~~~~~~G(x_2 - y_2, t- \tau_2)\delta(y_1 - y_2)\delta(\tau_1 - \tau_2) k(x_2)\bigg)\\
        &~~~~~~~~~~~~~~~~\cdot \exp\left(-i\int dx k(x) \phi(x) \right)
    \end{split} 
\end{equation}

which can be simplified to obtain 
\begin{equation}
    P_0[\phi, t] = \mathcal{N}^{-1} \int\mathcal{D}k~\exp\left( -\int dx_1 dx_2~ k(x_1)\frac{1}{2}D(x_1, x_2; t,t)~ k(x_2) - i\int dx~ k(x) \phi(x)\right) 
\end{equation}

Now performing Gaussian integration over \(k\), we obtain the probability density to be the distribution 
\begin{equation}
    P_0[\phi, t] = \mathcal{N}^{-1} \exp\left( -\frac{1}{2} \int dx_1 dx_2~\phi(x_1) D^{-1}(x_1, x_2; t,t) \phi(x_2)  \right)
    \label{eq:probDens}
\end{equation}

The propagator \(D\) can be explicitly calculated as
\begin{equation}
    \begin{split}
        D(x,y;t,t) &= 2\int dz\int_0^t d\tau \int dk_1 dk_2 ~~\mathrm{e}^{ik_1(x-z)-(k_1^2 + m^2)(t-\tau)} ~\mathrm{e}^{ik_2(y-z)-(k_2^2 + m^2)(t-\tau)}\\
        & = 2\int dk_1 dk_2 \int_0^t d\tau ~~\delta(k_1 + k_2) \exp\left(  ik_1x + ik_2y - (k_1^2 + k_2^2 + 2m^2)(t-\tau) \right)\\
        &=2\int dk\int_0^\tau ~\exp\left(ik(x - y) + 2(k^2 + m^2)(\tau - t)\right)\\
        &= \int dk~ \mathrm{e}^{ik(x - y)}\frac{1-\mathrm{e}^{-2t(k^2 + m^2)}}{ k^2 + m^2}
    \end{split}
\end{equation}
In the thermal equilbrium limit, the propagator reduces to 
\begin{equation}
    D(x,y) = \int dk~ \mathrm{e}^{ik(x - y)}\frac{1}{ k^2 + m^2} = \frac{-1}{\del^2 - m^2}
\end{equation}
which is the same as the propagator obtained from other quantization methods. From this, we see that the equation \ref{eq:probDens} in the thermal equilibrium limit gives the same measure, \(P \propto \exp(-S)\), as in the case of other quantization methods.

\section{Stochastic Quantization of Gauge Fields}
% Thus, the Euclidean Quantum Field Theory measures are solutions to which the diffusion equation given below relaxes 
% \begin{equation}
%     \pp P t  = \sum_i \pp{}{\varphi_i}\left(\left[ \hbar\pp{}{\varphi_i} - \pp{S}{\varphi_i} \right]P\right)
% \end{equation}
% Where the notation \(\varphi_i\) stands for \(\varphi(x)\) with \(x \equiv (x^0, x^1, \dots , x^{n-1})\) and the discrete sum over the index \(i\) actually represents the integration \(\int d^nx \) in the continuum.\\

Let us consider the simplest case of Abelian gauge fields, for which the Langevin equation (which we will call Parisi - Wu equations henceforth) reads
\begin{equation}
    \pp{A_\mu(x,t)}{t} = (\delta_{\mu\nu}\del^2 - \del_\mu \del_\nu)A_{\nu}(x,t) + \eta_\mu (x,t)
\end{equation}

The presence of the \(\displaystyle\frac{\del}{\del t}\) term for the fields \(A\) explicitly breaks the gauge invariance of the theory, meaning that one doesn't need to worry about fixing the gauge at all. In principle, one expects this explicit breaking of the gauge freedom to let us quantize the system without having to worry about fixing the gauge or the problems like Gribov Ambiguity that come with it, but this is not the case as we can see below. \\

Consider the Langevin equation for the gauge fields in the Fourier space, 
\begin{equation}
    \pp{A_\mu(k,t)}{t} = (\delta_{\mu\nu}k^2 - k_\mu k_\nu)A_{\nu}(k,t) + \eta_\mu (k,t)
\end{equation}

We can split the above equation into transverse and longitudinal parts, with
\begin{equation}
    \begin{split}
        A^T_\mu &= \left(\delta_{\mu\nu} - \frac{k_\mu k_\nu}{k^2}  \right)A_\nu = O_{\mu\nu}^T A_\nu\\
        A^L_\mu & = \frac{k_\mu k_\nu}{k^2} A_\nu = O_{\mu\nu}^L A_\nu\\
        \eta^T_\mu &=  \left(\delta_{\mu\nu} - \frac{k_\mu k_\nu}{k^2}  \right)\eta_\nu =O_{\mu\nu}^T\eta_\nu\\
        \eta^L_\mu & = \frac{k_\mu k_\nu}{k^2} \eta_\nu = O_{\mu\nu}^L \eta_\nu\\
    \end{split}
\end{equation}

This splits the Parisi-Wu equation as 
\begin{equation}
    \begin{split}
        \pp{A^T}{t} =& -k^2A^T + \eta^T\\
        \pp{A^L}{t} =&~ \eta^L
    \end{split}
\end{equation}

This equation has solutions
\begin{equation}
    \begin{split}
        A^T =& \exp(-k^2 t) A_0^T + \exp(-k^2 t)  \int_0^t \exp(k^2 \tau)\eta(\tau) d\tau\\
        A^L =& A_0^L + \int_0^t \exp(k^2 (t-\tau))\eta(\tau) d\tau
    \end{split}
\end{equation}
From this, we observe that the initial distribution in the transverse modes get dissipated at infinite time due to the presence of the damping term, while in the case of longitudinal mode, there is no damping term and hence the initial distribution persists even in equilibrium, i.e. there exists no stationary distribution in the transverse modes. This is directly the consequence of the gauge invariance. \\

To see how this affects calculations, let us look at the propagator for the gauge fields. 
\begin{equation}
    \begin{split}
        D_{\mu\nu}(k, t|k', t') =& \braket{A_\mu(k,t), A_\nu(k',t')}\\
        =& ~\delta^4(k+k') \frac{1}{k^2}  O^T_{\mu\nu}(\exp(k^2(t-t')) - \exp(-k^2(t+t'))) \\
        &~~~~~~~~~~~~+ 2t_<~\delta^4(k+k')\frac{k_\mu k_\nu}{k^2} + A_\mu(k,0)^LA_\nu(k',0)^L 
    \end{split}
\end{equation}

The translational invariance of the propagator requires that the propagator be proportional to \(\delta^4(k+k')\), which is broken by the last term in the above-derived propagator. One way to circumvent the problem would be to select the initial configuration of the longitudinal mode to be \(A_\mu(k, 0) = 0\). \\

But such a distribution is exceptional and therefore we consider a more general distribution that is symmetric around \(k=0\) as 
\begin{equation}
    A_{\mu}^L = \frac{k_\mu}{k^2}\phi(k)
\end{equation}
and for the propagator, we take another average over the distributions \(\phi(k)\). Knowing 
\begin{equation}
    \braket{\phi(k), \phi(k')}_\phi = -\alpha \delta^4(k+k')
\end{equation}
where \(\alpha\) is the width of \(\phi(k)\), we get the limit of the propagator as 
\begin{equation}
   \lim_{t=t'\to\infty} D_{\mu\nu}(k, t|k', t') =  {{\delta^{4}(k+k^{\prime})\left\{\frac1{k^{2}}\left(\delta_{\mu\nu}-(1-\alpha)\frac{k_{\mu}k_{\nu}}{k^{2}}\right)+2t\frac{k_{\mu}k_{\nu}}{k^{2}}\right\}}}
\end{equation}
which is nothing but the usual gauge fixed propagator, with a diverging term proportional to \(t\).\\

Therefore one can see that the stochastic quantization prescription allows one to quantize gauge fields and get the propagator of the theory without having to resort to gauge fixing explicitly, but as seen from the above calculations, the gauge freedom is reflected as a choice in the initial distribution for the longitudinal mode of the gauge fields, the width of the distribution \(\alpha\) playing the role of the gauge fixing parameter.\\

% Therefore, the gauge invariance reflects as a choice of the initial values of the longitudinal modes of the gauge configurations. This quantization scheme, although is consistant and is able to quantize the gauge theory without resorting to gauge fixing functional, it is termwise non-renormalisable. Since the propagators have terms proportional to \(t\), which are supposed to get cancelled in the summation of the different Feynman diagrams, a single graph of this is not renormalisable. \\

Another subtlety worth mentioning is that for gauge-invariant quantities, the term proportional to \(t\) in the equilibrium limit goes to zero. But the gauge-variant quantities diverge in the limit \(t\to \infty\). This is mainly due to the absence of drift force in the transverse part of Langevin's equation, meaning that the stationary distribution for the transverse part doesn't exist, and a particle undergoing Brownian motion in the configuration space will drift forever in this direction. Speaking in terms of forces, the action provides a force that constrains a particle to a gauge orbit in the configuration space, but the particle is free to move along the gauge orbit indefinitely. It is this free motion along the gauge orbits that causes the gauge-variant quantities to diverge in the equilibrium limit. \\

Therefore, for gauge-invariant quantities, the stochastic quantization prescription gives the same results as the regular quantization without having to resort to explicit gauge fixing procedures, but in the case of gauge-variant observables, the theory again leads to divergences. 

%%%%%%%%%%%%%%%%%%%%%%%%%
% Created on 03-10-2023 %
%                       %
% Author: Adithya A Rao %
%%%%%%%%%%%%%%%%%%%%%%%%%

\newpage
\chapter{Gauge Fixing without Fixing the Gauge}
\vspace{-45pt}
In this chapter, we discuss how the addition of a non-conservative drift force along the gauge orbit can provide a way to circumvent the problems associated with the stochastic quantization of gauge fields. We also discuss the properties of such a force and also discuss how such a formalism can be used to obtain the same effects as the restriction of the path integral to the Gribov region.  By avoiding the need for explicit gauge fixing and the introduction of ghost fields, stochastic quantization therefore offers a promising alternative for studying the non-perturbative aspects of gauge theories.

\section{Gauge Fixing as a Non-Conservative drift force}

As we saw from the previous sections, the divergences of gauge variant quantities at infinite time arise due to the free drift along the gauge orbits, where there are no damping forces to provide a stationary limit. Therefore, a way to circumvent this problem would be to introduce a drift force along the gauge orbits. In the proceeding section, we will see how such a force can be introduced and how the introduction of such a force leads naturally to the random walker being attracted to the Gribov region. \\

The presence of zeros of the Faddeev-Popov operator makes the friction force singular at the Gribov horizon. It is well known that if a particle undergoing Brownian diffusion in the presence of friction force encounters a boundary where the friction force is singular and repulsive enough, it will get reflected. Therefore, in this case, the introduced force being singular at the Gribov horizon implies that the Brownian motion in configuration space never leaves the Gribov region at all, and therefore, one can expect the stochastic quantization to inherently describe the same system as obtained by the regular quantization done by strictly restricting the path integral to the Gribov region. \\

% The introduced force can be seen to be singular at the Gribov horizon, and if the singular friction force is repulsive enough, a particle undergoing Brownian motion starting from inside the Gribov region will get reflected from the boundary every time it reaches it, never leaving the region. In such a case, the stochastic quantization inherently describes a theory where the path integral is already confined to the Gribov region.\\

\begin{figure}[h]
    \centering
    \begin{tikzpicture}[scale=0.75]
        \path[draw=black,line width=1px] (0,0) circle (4);
        \node at (5, 2.82) {Gribov Horizon};
        \node at (0,0) {Gribov Region};
        \filldraw [black] (-2,2) circle (2pt);
        \draw[thick] (-2,2) -- (-2.5,1.5) -- (-2.5, 1) -- (-3.25, 0.5) -- (-3, 0) -- (-3.5, -1) -- (-3.5, -1.75) -- (-2.75,  -2.25) -- (-2.25, -1.5) -- (-1.75, -2) -- (-0.5, -2.5) -- (0, -3.5) -- (1, -3.75) -- (2, -3) -- (1.5, -2);
        \draw[thick, ->] (1.5, -2)-- (2, -1) ;
    \end{tikzpicture}
    \caption{{The path of a particle undergoing Brownian motion in the configuration space in the presence of the singular friction force that is repulsive enough.}}
\end{figure}
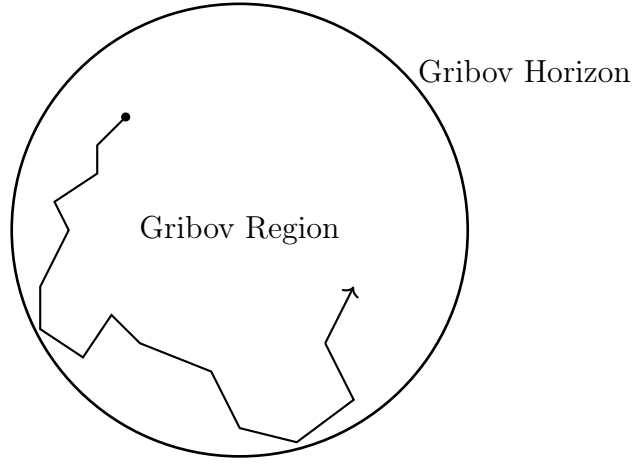

Consider a one-dimensional Brownian motion, with the probability density following the equation 
\begin{equation}
    \dot{P}(x,t) = \del (\del - f)P(x,t) = -L^*~P(x,t)
    \label{eq:dotPfromL}
\end{equation}
where \(f\) is the static drift force (all the constants have been taken to be 1).\\
The above equation has the solution 
\begin{equation}
    P(x,t) = e^{-L^*t} P(x,0) \equiv e^{-L^*t} P_0(x)
\end{equation}
Any functional that is dependent on \(x\), \(\Phi(x)\) has an expectation value given by 
\begin{equation}
    \begin{split}
        \langle\Phi \rangle_t &= \int dx ~\Phi(x)~P(x,t) = \int dx ~\Phi(x)[e^{-L^*t} P_0(x)]\\
        &= \int dx~\Phi(x)(1-L^*)P_0(x) ~~~~~~~~~~~~~~~~~~~~~~~~~~~~\text{(upto first order)}\\
        &=\int dx~ \Phi(x)(P_0 + \del^2P_0 - \del(fP_0))\\
        &\text{Performing integration by parts, and using } P_0(\pm \infty) = 0\text{~,~~we get}\\
        \langle\Phi \rangle_t &= \int dx~ (1 - \del^2 - f\del)\Phi(x)P_0(x) \equiv \int dx~[e^{-Lt}\Phi(x)] ~P_0(x)
    \end{split}
\end{equation}
Where \(L = -(\del + f)\del\).\\

Since as \(t\rightarrow \infty\), \(\langle\Phi\rangle_t\) relaxes to the \(\langle \Phi \rangle\) independent of \(P_0(x)\), so we can choose a trivial \(P_0(x) = \delta(x-x')\), and hence we arrive at 
\begin{equation}
    \langle \Phi\rangle = \lim_{t\rightarrow \infty} e^{-Lt}\Phi(x)
\end{equation}
Thus given an initial data \(\Phi(x,0)\), we can calculate \(\Phi(x,t)\) and the limit \(\displaystyle\lim_{t\rightarrow \infty} \Phi(x,t) = \langle\Phi\rangle\).\\

In the case of gauge fields, consider \(\Phi[A]\) which is a gauge invariant observable. Then the equation of motion becomes 
\begin{equation}
    \dot{\Phi}[A,t] = -L~\Phi[A,t]
\end{equation}
with 
\begin{equation}
    L = -\int d^Dx~ \left[  \DD{}{A(x)} ~-~\DD{S[A]}{A(x)}  \right]\DD{}{A(x)}
    \label{eq:LForGaugeInv}
\end{equation}
with \(S[A]\) being the classical action functional.\\

An infinitesimal gauge transformation is given as \(\delta A = D\omega = \del \omega + fA\omega\) (spacetime and color indices are understood).\\
Such a transformation changes the observable by 
\begin{equation}
    \delta\Phi[A] = \int d^dx D\omega~\DD{}{A(x)}\Phi[A]
\end{equation}
If \(\Phi\) is a gauge invarient observable, then \(\delta\Phi[A]\) should be zero.\\
Thus we can add a term \(\displaystyle\int  d^dx D\omega~\DD{}{A(x)}\) to eq (\ref{eq:LForGaugeInv}) without changing the value \(\langle \Phi\rangle_t\) provided that \(\Phi\) is gauge invarient observable.\\

This is equivalent to adding a term \(D\omega\) to the drift force \(f\). It is convenient to choose \(\omega\) to be a Lorentz scalar, and also a local function linear in \(A\). One such possibility is \(\omega \propto \del\cdot A\). The proportionality constant is the dimensionless gauge parameter \(\alpha^{-1}\), with \(\alpha=1\) being the Feynman Gauge and \(\alpha\rightarrow 0\) being the Landau gauge.   \\

Thus, we obtain the operator
\begin{equation}
        L_{\alpha} = -\int d^Dx~ \left[  \DD{}{A(x)} ~-~\DD{S[A]}{A(x)} ~-~\alpha^{-1}D\del A \right]\DD{}{A(x)}
\end{equation}
The term \(D\del A\) has two parts, \(\del^2 A\) and \(C = [A,\del A]\). The first term can be written as a conservative force, derived from the potential 
\begin{equation}
    S_G = \frac{1}{2}\int d^d x(\del A)^2
\end{equation}
while the second part cannot be written in this form, and hence is a non-conservative force. The \(S_G\) provides the necessary restoring force along the orbit, while the term \(C\) is a pure circulation \(\del . C = 0,~\del \times C \ne 0\) which in equilibrium produces a stationary circulating current. \\

The equation followed by the probability density becomes
\begin{equation}
    \dot{P_\alpha} = -L^*_\alpha P_\alpha
\end{equation}
and we see that the equilibrium distribution should follow \(L^*_\alpha P_{\alpha~E} = 0\). \\
This condition is not followed by the traditional Faddeev-Popov distribution 
\begin{equation}
    P_{\alpha,\mathrm{FP}}[A]=N\exp\left[-S-\alpha^{-1}S_{\mathrm{G}}\right]\mathrm{Det}\left(\partial^{\mu}D_{\mu}/\partial^{2}\right)
\end{equation}
Therefore the the \(P_{\alpha~E}\) is not the same as the FP distribution.

\section{Properties of the Gauge Fixing Force}
Let us look at the properties the introduced friction force has.
\begin{enumerate}
    \item \textbf{The force is zero (equilibrium point) when \(\del A = 0\): }\\
    This is pretty clear from the form of \(F_G = D \del A\) which vanishes when \(\del A = 0\). This implies that the gauge fixing hyperplane \(\del A = 0\) is made up of equilibrium configurations at which the friction force vanishes. 
    \item \textbf{It is a restoring force elsewhere:}\\
    Consider the Hilbert square norm \(||A||^2 = (A,A)\) where \((.,.)\) is the appropriate inner product. The time evolution of this in the presence of the gauge fixing force is given by (using \(\dd{A}{t} = F_G = \alpha^{-1}D\del A\))
    \begin{equation}
        \dd{||A||^2}{t} = (A, \dd{A}{t}) + (\dd{A}{t},A) = 2 ( A, \alpha^{-1} D\del A) 
    \end{equation}
    This is equal to (doing integration by parts)
    \begin{equation}
        \dd{||A||^2}{t} = -2\alpha^{-1} (DA, \del A) = -2\alpha^{-1}(\del A, \del A) = -2 \alpha^{-1} ||\del A||^2 \le 0
    \end{equation}
    Therefore, the drift force reduces the Hilbert square norm and hence is a restoring force. (provided \(\alpha>0\)). That is, a particle at any point on the configuration space that is not in the hyperplane \(\del A = 0\) is attracted towards the gauge fixing hyperplane and therefore drifts towards it along its corresponding gauge orbit. 
    \item \textbf{The equilibrium inside the Gribov horizon is stable and outside is unstable:}\\
    Consider 
    \begin{equation}
        \dd{||\del A||^2}{t} = \ -2\alpha^{-1}(\del A,-\del D_{A} \del A)
    \end{equation}
    At a point close to the equilibrium locus, \(A = A_0 + \epsilon A_1\), where \(A_0 \ni \del A_0 = 0\), since \(\del A_0 = 0\)the value of (upto lowest order in \(\epsilon\))
    \begin{equation} 
        \dd{||\del A||^2}{t} = -2\alpha^{-1}(\del A_1,-\del D_{A_0} \del A_1)\epsilon^2
    \end{equation}
    Thus, if \(A_0\) is inside the Gribov Horizon, i.e. \(-\del D_{A_0} > 0\), then the \(||\del A||^2\) is decreasing, and hence the point \(A_0\) is stable. \\
    If it is outside the Gribov horizon, \(||\del A||^2\) increases even for points close to \(A_0\) and hence the point \(A_0\) is unstable.\\
    Therefore the Gribov region \(\Omega\) is an attractor of the gauge fixing force.  
\end{enumerate}

From the properties of the friction force, one can see what happens to a random walker in the configuration space. 

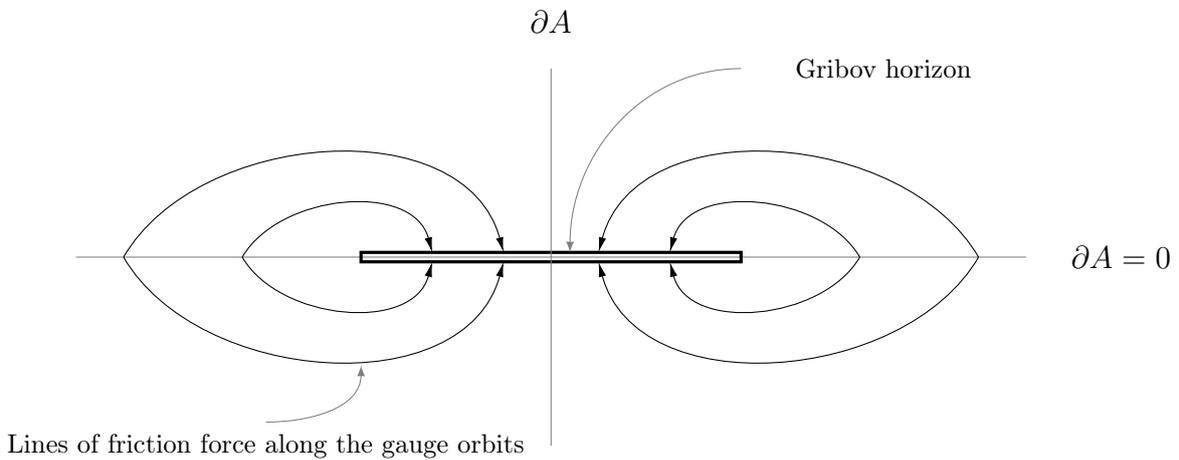
\begin{figure}[h]
    \centering
    \begin{tikzpicture}[scale=1.25]
        \node at (0, 2.5) {\(\del A\)};
        \node at (6, 0) {\(\del A = 0\)};
        \draw[light, gray] (-5, 0) -- (5,0);
        \draw[very thick] (-2, -0.05) rectangle (2,0.05);
        \draw[light, gray] (0, 2) -- (0,-2);
        \draw[{Latex[length=2mm,width=1mm]}-] (-1.25, -0.05) to[out = -105, in = -60] (-3.25, 0); 
        \draw[{Latex[length=2mm,width=1mm]}-] (-1.25, 0.05) to[out = 105, in = 60] (-3.25, 0); 
        \draw[{Latex[length=2mm,width=1mm]}-] (1.25, 0.05) to[out = 75, in = 120] (3.25, 0); 
        \draw[{Latex[length=2mm,width=1mm]}-] (1.25, -0.05) to[out = -75, in = -120] (3.25, 0); 
        
        \draw[{Latex[length=2mm,width=1mm]}-] (-0.5, -0.05) to[out = -105, in = -60] (-4.5, 0); 
        \draw[{Latex[length=2mm,width=1mm]}-] (-0.5, 0.05) to[out = 105, in = 60] (-4.5, 0); 
        \draw[{Latex[length=2mm,width=1mm]}-] (0.5, 0.05) to[out = 75, in = 120] (4.5, 0); 
        \draw[{Latex[length=2mm,width=1mm]}-] (0.5, -0.05) to[out = -75, in = -120] (4.5, 0); 

        \draw[gray, latex-] (0.2,0.05) to[out=90, in=180] (2,2);
        \node at (3.5, 2) {\footnotesize Gribov horizon}; 
        \draw[gray, latex-] (-2,-1.15) to[out=-90, in=0] (-3,-1.75);
        \node at (-3, -2) {\footnotesize Lines of friction force along the gauge orbits}; 
        \node at (0,-2.25) {~};
    \end{tikzpicture}
    \caption{{The gauge fixing frictional force confining the particle motion to the Gribov region}}
\end{figure}

The motion of the random walker in the configuration space in the presence of the frictional force along the gauge orbit can be classified into three categories based on its location. 
\begin{enumerate}
    \item \textbf{Outside the hyperplane} \(\mathbf{\del A = 0}\): \\
    The frictional force along the orbit the configuration belongs to will push the random walker toward the Gribov region.   
    \item \textbf{In the hyperplane and outside the Gribov region}: \\
    These points are equilibrium points of the force, and therefore they are unaffected by the friction force. However, the thermal fluctuations due to the noise force will push the walker away from the hyperplane, at which point it is pushed away from the hyperplane along the orbit towards the Gribov region.
    \item \textbf{In the hyperplane and inside the Gribov region}:\\
    At these points, the thermal fluctuations cannot push the walker in the transverse direction but can push the walker along the hyperplane in the direction that takes it outside the Gribov region. In the case the force at the horizon is repulsive enough, the walker is reflected and is therefore confined to the Gribov region. Even in the case the force is not repulsive enough, the walker goes out of the Gribov region to the region where the force stops being attractive. It again gets some transverse component due to thermal fluctuations, and again gets pushed along gauge orbit towards the Gribov region. 
\end{enumerate}

Therefore we see that a random walker starting at any position in the configuration space drifts towards the Gribov region, and therefore there is no random drifting to infinity like in the case with no friction force. This procedure eliminates the problems with stochastic quantization prescription, while also allowing us to quantize the gauge fields without having to explicitly perform the gauge fixing, while also circumventing the Gribov problem. 

%%%%%%%%%%%%%%%%%%%%%%%%%
% Created on 03-10-2023 %
%                       %
% Author: Adithya A Rao %
%%%%%%%%%%%%%%%%%%%%%%%%%

\newpage
\part{Concluding Remarks}

\newpage
% \thispagestyle{plain}
% \vspace{30pt}

% \chapter{Future scope}
% \section*{Conclusion \dots}

The Gribov problem in quantum Yang-Mills theories arises due to the presence of multiple solutions to the gauge fixing functional. This leads to ill-defined propagators, and incorrect calculations of observables due to overcounting of the gauge copies, along with a complete breaking of the introduction of the Faddeev-Popov ghosts. Therefore, the Gribov ambiguity presents a formidable problem to the non-perturbative studies of the quantum Yang-Mills theories. One could argue that the gaps in our understanding of the non-perturbative aspects of these theories are due to the non-trivial structure of the configuration space which makes the analysis very difficult, and there is a need to study analytically the configuration space of the theory to obtain deeper understandings into the non-perturbative regimes. \\

Restricting the path integral to the fundamental modular region is a proposed resolution, but so far no attempts have succeeded in imposing such a restriction. The semi-classical restriction to the Gribov region does indeed circumvent the problem to an extent. but it is an approximation that does not capture all the information about the theory. Some authors have suggested that the restriction is expected to lead to the generation of the mass gap in the theory and possibly confinement as seen in a rudimentary version of calculations, but the exact mechanism behind this is largely unknown. Further, Zwanziger's attempts \cite{zwanzigerLocalRenormalizableAction1989, zwanzigerActionGribovHorizon1989} at resolving this problem have led to a non-local action formalism for the quantum Yang-Mills, the discussion of which is beyond the scope of this thesis. \\

The stochastic quantization formalism, which quantizes the gauge fields as a Brownian motion in the configuration space, provides an alternative quantization that avoids the Gribov problem. This formalism quantizes the gauge theory by the use of Langevin's equation which has an operator \(\del/\del t\) that breaks the gauge invariance of the theory and allows us to quantize it. Further, one can circumvent the problem of the absence of a stationary distribution in the transverse modes for the gauge fields by introducing a restoring force along the Gribov orbit. The restoring force confines a random walker's probability density to the Gribov region, therefore reproducing the same effect as that of restriction of the path integral to the Gribov region. Further, if the restoring force is repulsive enough at the Gribov horizon, the random walker will never leave the Gribov region at all, exactly reproducing the effects of the restricted path integral. However, the exact behavior of the restoring force is unknown, and therefore the behavior of the walker at the boundaries cannot be determined. Further Baulieu and Zwanziger have proved the equivalence of the stochastic quantization prescription to the Faddeev-Popov prescription \cite{baulieuEquivalenceStochasticQuantization1981} implying the equivalence of the calculations done via the stochastic methods. Research is being conducted by researchers like Ajay Chandra \cite{chandraStochasticQuantisationYangMillsHiggs2022, chandraLangevinDynamic2D2022}, Massimiliano Gubinelli \cite{gubinelliPDEConstructionEuclidean2021}, Arthur Jaffe \cite{jaffeQuantumFieldsStochastic2015}, and others towards a mathematically rigorous quantization of fields via stochastic quantization. \\

In summary, the Gribov problem presents a significant obstacle to the non-perturbative study of quantum Yang-Mills theories, and while various approaches have been proposed to address this issue, a complete understanding remains elusive. The stochastic quantization formalism offers a promising alternative by circumventing the gauge-fixing procedure and introducing a restoring force along the Gribov orbit, but further investigation is needed to fully understand the behavior and the equilibrium limits of the system.  

\newpage
\addcontentsline{toc}{chapter}{List of Figures}
\listoffigures

%%%%%%%%%%%%%%%%%%%%%%%%%%%%%%%%%%%%%%%%%%%%%%%%%%%%%%%%%%%%%%%%%%%%%%%%%%%%%

% Bibliography

\nocite{*}
\newpage
\bibliographystyle{ieeetr}
\addcontentsline{toc}{chapter}{Bibliography}

\bibliography{references.bib}

% \newpage
% \thispagestyle{empty}
% \addtocontents{toc}{\protect \contentsline {chapter}{Certificate}{iii}{}}
% \includepdf[pages={1}]{preamble/scanned.pdf}
% \includepdf[pages={2,3,4}]{preamble/scanned.pdf}
% \input{preamble/plagiarism_declaration.tex}

\end{document}